\documentclass[preprint,12pt]{elsarticle}

\usepackage{graphicx} 
\usepackage{amsmath}  
\usepackage{placeins} 
\usepackage{booktabs} 
\usepackage{longtable} 
\usepackage{amssymb}  
\usepackage{hyperref}
\usepackage{adjustbox}
\usepackage{siunitx}
\usepackage{tabularx}
\usepackage{pdflscape}
\usepackage{tikz}
\usetikzlibrary{shapes.geometric, arrows.meta, positioning, fit, shadows, decorations.pathreplacing}
\usepackage{xcolor}

\journal{Computers and Chemical Engineering}

\begin{document}

\newcommand{\inb}[1]{\left(#1\right)}
\newcommand{\bm}[1]{\mathbf{#1}}
\newcommand{\colored}[2]{\textcolor{#1}{#2}}
\newcommand{\emptyline}{\vspace{12pt}}
\newcommand{\safespace}[1]{#1~}

\newcommand{\vectorized}[1]{#1}  

\newcommand{\propertyphase}[2]{#1^{#2}}
\newcommand{\propertyindex}[2]{#1_{#2}}
\newcommand{\propertyphaseindex}[3]{\propertyindex{\propertyphase{#1}{#2}}{#3}}

\newcommand{\best}{*}
\newcommand{\mynum}{N}
\newcommand{\column}{Co}
\newcommand{\component}{C}
\newcommand{\spl}{s}
\newcommand{\refluxratio}{RR}
\newcommand{\boilupratio}{BR}
\newcommand{\Acetone}{Ac}


\newcommand{\ones}{\vectorized{1}}

\newcommand{\figureref}[1]{Fig.~\ref{#1}}
\newcommand{\sectionref}[1]{Sec.~\ref{#1}}
\newcommand{\tabref}[1]{Tab.~\ref{#1}}
\newcommand{\appendixref}[1]{\ref{#1}}


\newcommand{\vapor}{V}
\newcommand{\liquid}{L}
\newcommand{\critical}{C}
\newcommand{\saturated}{\mathcal{S}}
\newcommand{\saturatedvapor}{\saturated \vapor}
\newcommand{\vaporization}{\Delta \vapor}

\newcommand{\modelfluid}{MF}
\newcommand{\margules}{M}
\newcommand{\simplifiedantoine}{SA}
\newcommand{\extendedantoine}{EA}
\newcommand{\entrainer}{E}
\newcommand{\process}{P}

\newcommand{\rigorous}{Rig}

\newcommand{\total}{tot}
\newcommand{\reboiler}{Reb}
\newcommand{\condenser}{Con}
\newcommand{\product}{Pro}
\newcommand{\hypothetical}{Hypo}
\newcommand{\candidate}{Can}
\newcommand{\recycle}{Rec}
\newcommand{\feed}{Fe}
\newcommand{\bottom}{Bo}
\newcommand{\distillate}{Di}
\newcommand{\flowsheet}{Fs}
\newcommand{\abovefeed}{AF}
\newcommand{\belowfeed}{BF}
\newcommand{\stages}{S}

\newcommand{\liquidmolarfraction}{\ell}
\newcommand{\liquidmolarfractions}{\vectorized{\liquidmolarfraction}}
\newcommand{\vapormolarfraction}{v}
\newcommand{\vapormolarfractions}{\vectorized{\vapormolarfraction}}
\newcommand{\vaporflow}{V}
\newcommand{\liquidflow}{L}
\newcommand{\temperature}{T}
\newcommand{\pressure}{p}
\newcommand{\molarenthalpy}{h}
\newcommand{\feature}{x}
\newcommand{\featurespace}{\mathcal{X}}
\newcommand{\parameter}{\theta}
\newcommand{\activitycoefficient}{\gamma}
\newcommand{\heatduty}{\dot{Q}}
\newcommand{\reboilerduty}{\heatduty^{\reboiler}}
\newcommand{\totalreboilerduty}{\heatduty^{\reboiler, \total}}
\newcommand{\condenserduty}{\heatduty^{\condenser}}

\newcommand{\model}{f}
\newcommand{\loss}{\mathcal{L}}
\newcommand{\constraint}{\model}
\newcommand{\jacobian}{J}


\newcommand{\nrtlparameters}{\parameter^{\text{NRTL}}}
\newcommand{\extendedantoineparameters}{\parameter^{\text{\extendedantoine}}}
\newcommand{\modelfluidfeature}{\feature^{\text{\modelfluid}}}
\newcommand{\modelfluidfeatures}{\vectorized{\modelfluidfeature}}
\newcommand{\modelfluidparameter}{\parameter^{\modelfluid}}
\newcommand{\modelfluidparameters}{\vectorized{\modelfluidparameter}}

\newcommand{\liquidmolarenthalpy}{\propertyphase{\molarenthalpy}{\liquid}}
\newcommand{\vapormolarenthalpy}{\propertyphase{\molarenthalpy}{\vapor}}
\newcommand{\vaporizationenthalpy}{\propertyphase{\molarenthalpy}{\vaporization}}

\newcommand{\saturatedvaportemperature}{\propertyphase{\temperature}{\saturatedvapor}}
\newcommand{\saturatedvaportemperatureof}[1]{\propertyindex{\saturatedvaportemperature}{#1}}
\newcommand{\liquidmolarfractionof}[1]{\propertyindex{\liquidmolarfraction}{#1}}
\newcommand{\vapormolarfractionof}[1]{\propertyindex{\vapormolarfraction}{#1}}
\newcommand{\vaporizationenthalpyof}[1]{\propertyindex{\vaporizationenthalpy}{#1}}

\newcommand{\derivativefeatureof}[1]{\frac{\partial \vapormolarfractionof{#1}}{\partial \liquidmolarfractionof{#1}}}
\newcommand{\derivativefeatureatinfdilution}[2]{\derivativefeatureof{#1}\vert_{#2}}
\newcommand{\activitycoefficientatinfinitedilutionin}[2]{\activitycoefficient_{#1}\vert_{#2}}
\newcommand{\saturatedvaporpressurevariantaofat}[2]{\pressure_{#1, \vapor}^{\saturated, \rigorous}\inb{#2, \extendedantoineparameters}}
\newcommand{\activitycoefficientvariantaofat}[1]{\activitycoefficient_{#1}^{\rigorous}\inb{\liquidmolarfractionof{#1}=0, \temperature\inb{\liquidmolarfractionof{#1}=0}}}

\newcommand{\barunit}{\text{bar}}
\newcommand{\molunit}{\text{mol}}

\tikzstyle{stagenumber} = [font=\bfseries] 

\newcommand{\normalstage}[2]{
    \begin{scope}[shift={(0,#2)}]  
        \draw (0, 0) rectangle (2, 0.5);
        \draw [->] (0.5, 0.) -- (0.5, -0.5);
        \draw [->] (1.5, 0.5) -- (1.5, 1.);
        \node[stagenumber] at (0.5, 0.25) {$#1$};
        \node[stagenumber] at (1.5, 0.25) {$\temperature^{({#1})}$};
        \node[stagenumber] at (3.5, 0.75) {$\vaporflow^{(\stages^{({#1})})}$, $\vapormolarfractions^{(\stages^{({#1})})}$};
        \node[stagenumber] at (-1.5, -0.25) {$\liquidflow^{(\stages^{({#1})})}$, $\liquidmolarfractions^{(\stages^{({#1})})}$};
    \end{scope}
}

\newcommand{\bottomstage}[2]{
    \begin{scope}[shift={(0,#2)}]  
        \draw (0.5, 0.44) circle (0.05);

        \draw [->] (0.5, 0.40) -- (0.5, -0.75);
        \draw [->] (0.54, 0.44) -- (1.5, 0.44) -- (1.5, 1.0);
        
        \draw (1.0, 0.25) circle (0.35);
        \draw [-] (0.75, -0.35) -- (0.75, 0.35);
        \draw [-] (1.25, -0.35) -- (1.25, 0.35);
        \draw [-] (0.75, 0.35) -- (1., 0.15);
        \draw [-] (1., 0.15) -- (1.25, 0.35);

        \node[stagenumber] at (1.1, -0.6) {$\reboilerduty$};
        \node[stagenumber] at (0.1, 0.3) {$\boilupratio$};
        \node[stagenumber] at (3, 0.75) {$\vaporflow^{({#1})}$, $\vapormolarfraction^{({#1})}$, $\temperature^{({#1})}$};
        \node[stagenumber] at (-1, -0.75) {$\liquidflow^{\bottom}$, $\liquidmolarfraction^{\bottom}$, $\temperature^{\bottom}$};
    \end{scope}
}

\newcommand{\distillatestage}[2]{
    \begin{scope}[shift={(0,#2)}]  
        \draw (1.5, 0.05) circle (0.05);

        \draw [->] (1.5, 0.1) -- (1.5, 0.8);
        \draw [->] (1.45, 0.05) -- (0.5, 0.05) -- (0.5, -0.5);

        \draw (1.0, 0.25) circle (0.35);
        \draw [-] (0.75, 0.15) -- (0.75, 0.85);
        \draw [-] (1.25, 0.15) -- (1.25, 0.85);
        \draw [-] (0.75, 0.15) -- (1., 0.35);
        \draw [-] (1., 0.35) -- (1.25, 0.15);

        \node[stagenumber] at (1.0, 1.2) {$\condenserduty$};
        \node[stagenumber] at (1.9, 0.2) {$\refluxratio$};

        \node[stagenumber] at (-2.2, -0.25) {$\liquidflow^{({#1})}$, $\liquidmolarfraction^{({#1})}$, $\temperature^{({#1})}$};
        \node[stagenumber] at (2.6, 1) {$\liquidflow^{\distillate}$, $\liquidmolarfraction^{\distillate}$, $\temperature^{\distillate}$};
    \end{scope}
}

\newcommand{\feedstage}[2]{
    \begin{scope}[shift={(0,#2)}]  
        \draw (0, 0) rectangle (2, 0.5);
        \draw [->] (0.5, 0.) -- (0.5, -0.5);
        \draw [->] (1.5, 0.5) -- (1.5, 1.);
        \draw [->] (-1., 0.25) -- (0., 0.25);
        \node[stagenumber] at (0.5, 0.25) {$#1$};
        \node[stagenumber] at (1.5, 0.25) {$\temperature^{({#1})}$};
        \node[stagenumber] at (3.5, 0.75) {$\vaporflow^{({#1})}$, $\vapormolarfraction^{({#1})}$};
        \node[stagenumber] at (-1, -0.4) {$\liquidflow^{({#1})}$, $\liquidmolarfraction^{({#1})}$};
        \node[stagenumber] at (-2.5, 0.25) {$\liquidflow_{\feed}$, $\liquidmolarfraction_{\feed}$, $\temperature_{\feed}$};
    \end{scope}
}

\begin{frontmatter}

    \title{A Machine Learning-Fueled Modelfluid for Flowsheet Optimization}

    \author[1]{Martin Bubel\corref{cor1}}
    \author[1]{Tobias Seidel}
    \author[1]{Michael Bortz}
    \cortext[cor1]{Corresponding author. Email: martin.bubel@itwm.fraunhofer.de}
    \affiliation[1]{
        organization={Department of Optimization, Fraunhofer Institute for Industrial Mathematics},
        addressline={Fraunhofer-Platz 1},
        city={Kaiserslautern},
        postcode={D-67663},
        state={Rheinland-Pfalz},
        country={Germany}
    }

    \begin{abstract}
        Process optimization in chemical engineering may be hindered by the limited availability of reliable thermodynamic data for fluid mixtures.
        Remarkable progress is being made in predicting thermodynamic mixture properties by machine learning techniques.
        The vast information provided by these prediction methods enables new possibilities in process optimization.
        This work introduces a novel modelfluid representation that is designed to seamlessly integrate these ML-predicted data directly into flowsheet optimization.
        Tailored for distillation, our approach is built on physically interpretable and continuous features derived from core vapor liquid equilibrium phenomena.
        This ensures compatibility with existing simulation tools and gradient-based optimization.
        We demonstrate the power and accuracy of this ML-fueled modelfluid by applying it to the problem of entrainer selection for an azeotropic separation.
        The results show that our framework successfully identifies optimal, thermodynamically consistent entrainers with high fidelity compared to conventional models.
        Ultimately, this work provides a practical pathway to incorporate large-scale property prediction into efficient process design and optimization, overcoming the limitations of both traditional thermodynamic models and complex molecular-based equations of state.
    \end{abstract}




\end{frontmatter}

\section*{CRediT authorship contribution statement}
Author A: Writing - original draft; Formal analysis; Investigation; Software; Visualization; Methodology; Data curation; Validation.

Author B: Conceptualization; Formal analysis; Validation; Investigation; Supervision; Writing - review \& editing.

Author C: Conceptualization; Formal analysis; Supervision; Writing - review \& editing.

\section*{Funding}
No funding was received for this work.

\section{List of Operators, Variables, and Abbreviations}
\begin{tabular}{ll}
    \textbf{Operator} & \textbf{Meaning} \\
    $\hat{a}$ & Estimate of $a$ \\
    $a^*$ & Optimal value of $a$ according to some objective function \\
    $a_{i}$ & Index $i$ of $a$ \\
    $a^{(k)}$ & Stage $k$ of $a$ (regarding distillation column stages) \\
    $a^{\text{phase}}$ & Property $a$ in given phase \\
    $a^{\text{type}}$ & Property $a$ of given type \\
\end{tabular}

\emptyline
\noindent
\begin{tabular}{ll}
    \textbf{Variable} & \textbf{Meaning} \\
    $\ell$ & Liquid molar fraction \\
    $v$ & Vapor molar fraction \\
    $\liquidmolarfraction$ & Liquid molar fraction(s) \\
    $\vapormolarfraction$ & Vapor molar fraction(s) \\
    $\liquidflow$ & Liquid molar flow rate \\
    $\vaporflow$ & Vapor molar flow rate \\
    $\temperature$ & Temperature \\
    $\pressure$ & Pressure \\
    $\molarenthalpy$ & Molar enthalpy \\
    $\activitycoefficient$ & Activity coefficient \\
    $\feature$ & Feature \\
    $\featurespace$ & Feature space \\
    $\parameter$ & Parameter \\
    $\reboilerduty$ & Reboiler heat duty \\
    $\condenserduty$ & Condenser heat duty \\
    $\jacobian$ & Jacobian matrix \\
    $\model$ & Model function \\
    $\refluxratio$ & Reflux ratio \\
    $\boilupratio$ & Boil-up ratio \\
\end{tabular}

All variables are used in SI-units unless otherwise stated.

\emptyline
\noindent
\begin{tabular}{ll}
    \textbf{Abbreviation} & \textbf{Meaning} \\
    $\modelfluid$ & Modelfluid \\
    $\margules$ & Margules model \\
    $\simplifiedantoine$ & Simplified Antoine equation \\
    $\extendedantoine$ & Extended Antoine equation \\
    $\entrainer$ & Entrainer \\
    $\process$ & Process \\
    $\rigorous$ & Rigorous (model/parameter) \\
    $\total$ & Total \\
    $\reboiler$ & Reboiler \\
    $\condenser$ & Condenser \\
    $\product$ & Product \\
    $\hypothetical$ & Hypothetical \\
    $\candidate$ & Candidate \\
    $\recycle$ & Recycle \\
    $\feed$ & Feed \\
    $\bottom$ & Bottom product \\
    $\distillate$ & Distillate product \\
    $\flowsheet$ & Flowsheet \\
    $\abovefeed$ & Above feed stage(s) \\
    $\belowfeed$ & Below feed stage(s) \\
    $\stages$ & Stages \\
    ML & Machine Learning \\
    VLE & Vapor-Liquid Equilibrium \\
    PCP & Pure Component Property \\
    w.r.t. & With respect to \\
\end{tabular}

\section{Introduction \label{sec:introduction}}

The chemical industry faces mounting pressure from economic and environmental factors to enhance process efficiency. While advanced computational methods hold promise, their application in large-scale chemical engineering design studies may be hampered by a critical bottleneck: the scarcity of reliable data, especially for fluid mixtures. This challenge is particularly acute in computer-aided molecular and process design (CAM(P)D), where the optimization of process fluids is a key objective.

At the heart of process fluid optimization lies the concept of a \textit{modelfluid}: a parametric representation used to predict the thermodynamic and physical properties of a fluid mixture.
An effective modelfluid enables the optimization of the fluid itself, for instance, by selecting the optimal solvent or entrainer for a separation process.
This is where the modelfluid representation can make a difference, as most of the existing ones have not been developed with a focus on optimization-friendliness.

\paragraph{Requirements for an optimization-friendly modelfluid}
Based on the needs of process fluid optimization, we propose that an ideal modelfluid representation should satisfy four key requirements:
\begin{description}
    \item[R1: Availability.] Features must be readily available or easily predictable, particularly for mixtures where experimental data is scarce.
    
    \item[R2: Integrability.] The representation must be straightforward to integrate into existing process simulation software.
    
    \item[R3: Interpretability.] Features should be physically meaningful, ideally with known units and bounds, to allow for a clear interpretation of optimization results.
    
    \item[R4: Continuity.] The feature space must be continuous to ensure compatibility with gradient-based optimization algorithms and enable smooth exploration.
\end{description}

\paragraph{Limitations of existing Modelfluid Representations}
Traditional thermodynamic models (e.g., NRTL, UNIFAC) satisfy R2 and R4, as they are standard in simulators.
Also, they generalize well over conditions such as temperature and composition, as noted in \cite{Jirasek2022}.
However, their parameters often are of empirical nature and thus not physically interpretable, failing \textbf{R3}, so generalization over systems may be limited \cite{Jirasek2022}.
More critically, reliable parameter sets for a wide range of mixtures are scarce, violating \textbf{R1} \cite{Jirasek2020}.
Advanced, molecular-based modelfluids, such as those derived from the Statistical Associating Fluid Theory (SAFT), are a prominent modelfluid choice in the literature on process fluid optimization (see Section~\ref{sec:literaturereview}).
While powerful, SAFT-based approaches present their own challenges -- in particular w.r.t. \textbf{R3}, as noted e.g. in Prausnitz et al.~\cite[p. 391]{Prausnitz1998}.
\paragraph{Overcoming the Data Scarcity Problem}
Recently, machine learning and other advanced computational methods have emerged to directly address the challenge of limited mixture property data.
A growing body of work focuses on predicting key thermodynamic properties, such as activity coefficients at infinite dilution, for vast numbers of unmeasured systems \cite{Jirasek2020,Rittig2023,Specht2024}.
However, a gap remains in fully leveraging these large-scale predicted databases for process optimization.
A promising step in this direction was taken by Medina et al.~\cite{Medina2025}, who used prediction-based activity coefficients to parameterize a Margules model for VLE prediction.
This highlights the potential for custom modelfluids designed specifically to capitalize on the outputs of modern prediction methods.

\paragraph{Contribution of this Work}
In this work, we introduce a novel modelfluid representation tailored for the optimization of distillation-based processes.
Our modelfluid is built upon physically meaningful features derived from vapor liquid equilibrium (VLE) characteristics, such as singular points and relative volatilities.
This design ensures that all four requirements are met: the features are \textit{available} (since the framework can be parameterized using data from experiments, established thermodynamic models, or sourced from new prediction methods to overcome data scarcity), \textit{integrable}, \textit{interpretable}, and create a \textit{continuous} optimization space.
We demonstrate the utility of this modelfluid by applying it to the problem of entrainer selection, providing a guided exploration of the candidate space.

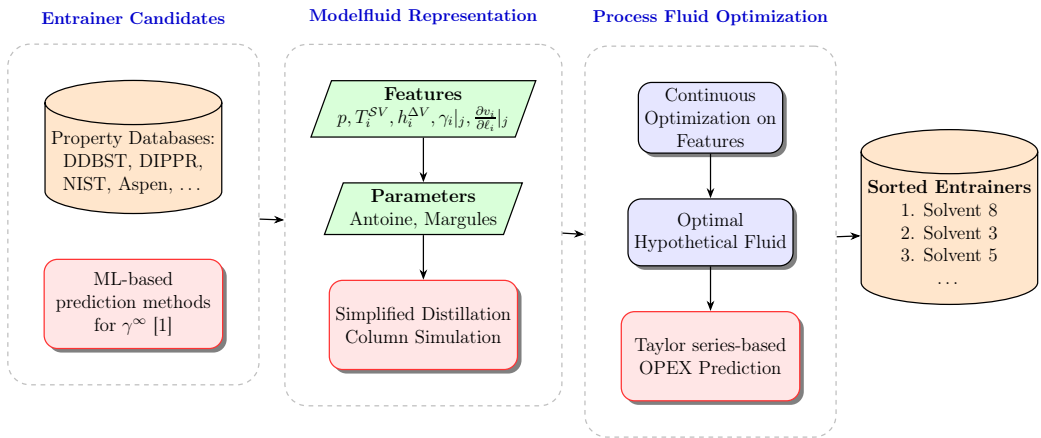
\begin{figure*}[t!]

\centering
\begin{adjustbox}{width=\textwidth, center}  
\begin{tikzpicture}[
    node distance=10mm and 15mm,
    process/.style={
        rectangle, 
        rounded corners=3mm, 
        draw, 
        thick, 
        fill=blue!10, 
        align=center,
        minimum height=15mm, 
        minimum width=30mm,
        drop shadow
    },
    data/.style={
        trapezium, 
        trapezium left angle=70, 
        trapezium right angle=110,
        draw, 
        thick, 
        fill=green!15, 
        align=center, 
        minimum height=10mm
    },
    database/.style={
        cylinder, 
        shape border rotate=90, 
        aspect=0.25, 
        draw, 
        thick, 
        fill=orange!20,
        align=center, 
        minimum height=15mm, 
        minimum width=25mm
    },
    digitaltwin/.style={
        rectangle, 
        rounded corners=3mm, 
        draw=red!80, 
        thick, 
        fill=red!10, 
        align=center,
        minimum height=20mm, 
        minimum width=40mm,
        drop shadow
    },
    arrow/.style={
            -Stealth,
            thick
        },
    label/.style={
        text=blue!80!black, 
        font=\small\bfseries
    },
    fitbox/.style={
        rectangle,
        rounded corners=5mm,
        draw=gray!50,
        thick,
        dashed
    }
]

\node (pcp_databases) [database] {Property Databases: \\ DDBST, DIPPR, \\ NIST, Aspen, \dots};
\node (ml_mc) [digitaltwin, below=of pcp_databases] {ML-based \\ prediction methods \\ for $\gamma^\infty$ \cite{Jirasek2020}};
\node (phase1_box) [fitbox, fit=(pcp_databases) (ml_mc), inner sep=8mm] {};
\node[label, above=3mm of phase1_box.north] {Entrainer Candidates};

\node (features)  [data, right=of pcp_databases, xshift=7mm, yshift=12mm] {\textbf{Features} \\ $\pressure, \saturatedvaportemperatureof{i}, \vaporizationenthalpyof{i}, \activitycoefficientatinfinitedilutionin{i}{j}, \derivativefeatureatinfdilution{i}{j}$};
\node (parameters) [data, below=of features] {\textbf{Parameters} \\ Antoine, Margules};
\node (simplifiedcolumn) [digitaltwin, below=of parameters] {Simplified Distillation \\ Column Simulation};

\draw[arrow] (features) -- (parameters);
\draw[arrow] (parameters) -- (simplifiedcolumn);

\node (phase2_box) [fitbox, fit=(features) (parameters) (simplifiedcolumn), inner sep=8mm] {};
\node[label, above=3mm of phase2_box.north] {Modelfluid Representation};

\draw[arrow] (phase1_box) -- (phase2_box);

\node (pfo) [process, right=of features, xshift=10mm, yshift=-2mm] {Continuous \\ Optimization on \\ Features};
\node (hypofluid) [process, below=of pfo] {Optimal \\ Hypothetical Fluid};
\node (opexprediction) [digitaltwin, below=of hypofluid] {Taylor series-based \\ OPEX Prediction};

\draw[arrow] (pfo) -- (hypofluid);
\draw[arrow] (hypofluid) -- (opexprediction);

\node (phase3_box) [fitbox, fit=(pfo) (hypofluid) (opexprediction), inner sep=8mm] {};
\node[label, above=3mm of phase3_box.north] {Process Fluid Optimization};

\draw[arrow] (phase2_box) -- (phase3_box);

\node (sortedentrainers) [database, right=of hypofluid] {\textbf{Sorted Entrainers} \\ 1. Solvent 8 \\ 2. Solvent 3 \\ 3. Solvent 5 \\ \dots};
\draw[arrow] (phase3_box) -- (sortedentrainers);

\end{tikzpicture}
\end{adjustbox}
\caption{
    Starting by a set of \textbf{entrainer candidates}, data from both literature databases and ML-based prediction methods are used to obtain the \textbf{modelfluid features}.
    A \textbf{process fluid optimization} based on those features is combined with an objective approximation method to yield a list of \textbf{sorted entrainer candidates}.
    This workflow demonstrates the embedding of the proposed modelfluid representation into a process fluid optimization task.
}
    \label{fig:methodstack}
\end{figure*}

In \figureref{fig:methodstack} we provide a schematic overview of the proposed modelfluid representation and its embedding into a process optimization task.
The methods and steps in \figureref{fig:methodstack} are introduced, discussed, and analyzed throughout this work.
%

This paper is organized as follows: Section~\ref{sec:literaturereview} reviews the field of process fluid optimization. Section~\ref{sec:modelfluid} introduces our novel modelfluid representation in detail. Section~\ref{sec:entrainersearch} demonstrates its application in an entrainer selection case study.
Finally, Section~\ref{sec:conclusion} summarizes our findings and concludes the work.

\section{Literature Review \label{sec:literaturereview}}
The field of process fluid optimization (PFO), often used interchangeably with computer-aided molecular design (CAMD), has seen significant evolution over the past three decades.
This section provides a focused overview of this progression, highlighting the key developments in molecular representation, property prediction, and optimization strategies.
The central goal is to contextualize the novel modelfluid representation we develop for distillation-based separation processes.

A typical PFO framework is built upon three core components: the modelfluid representation that parameterizes a molecule, the property prediction models that link this representation to physical behavior, and the optimization algorithm that searches the design space for optimal candidates \cite{Bestwick2023}.

The evolution of PFO can be understood through the advancements in each of these areas.
Early PFO research, from the 1990s and early 2000s, was dominated by discrete molecular representations.
These approaches often described molecules using structural groups and topological indices, leading naturally to integer programming or enumeration-based optimization problems \cite{Odele1993,Stefanis1996,Churi1996,Sinha1999,Marcoulaki2000,Wang2002A,Wang2002B}.
Given the non-convex nature of these problems, global optimization techniques, particularly genetic algorithms, were frequently employed \cite{Sinha1999, Marcoulaki2000,Wang2002B,Sinha2003,Hugo2004,Lehmann2004,Papadopoulos2005,Geroulis2023}.
However, a key challenge was the combinatorial explosion of the search space, which prompted the development of methods for search space reduction, such as applying structural constraints \cite{Churi1996}, molecular clustering \cite{Papadopoulos2006B}, and other creative heuristics \cite{Harper1999}.
Despite their combinatorial complexity and integer programming, the idea of optimizing on the molecular structure remains relevant in the field computer-aided molecular (process) design (CAM(P)D) as can be seen by the reviews of \cite{Papadopoulos2018,Iftakher2023,Gertig2020,Austin2016}.
While powerful, representations based on topological indices have a significant drawback: they "\textit{do not necessarily have a causal relationship with the correlated property}" \cite{Raman1998}, which limits their physical interpretability.

The introduction of fully continuous optimization using molecular descriptors marked a significant shift in PFO.
While earlier deterministic approaches used mixed-integer formulations \cite{Sheldon2006}, the work of \cite{Bardow2010} was the first to formulate the PFO problem in a fully continuous space.
This paradigm shift was enabled by using parameters from sophisticated equations of state, primarily SAFT-based models, as the continuous design variables.
This powerful approach has since been extended and refined for various applications, including carbon capture and solvent design \cite{Stavrou2014,Lampe2015,Stavrou2023,Mairhofer2023}.
While the nature of continuous optimization is less challenging than combinatorial mixed-integer problems, the continuous solution may not correspond to a real molecule which necessitates a subsequent discretization step \cite{Bardow2010}.
Despite their success, continuous PFO methods often require expert knowledge to pre-define the chemical search space \cite{Stavrou2023}, and their underlying parameters can lack a direct, intuitive link to macroscopic fluid properties.

Recent advances in the development of surrogate models that generalize over multiple mixtures, as found in \cite{Sun2024}, provide further opportunities in process fluid optimization due to the computational efficiency of surrogate models and the usage of molecular descriptors as input features.
Also, the usage of VLE-based surrogate models, as found in \cite{Sun2023,Sethi2025} can be used to alleviate the computational cost of solving MESH equation systems in rigorous distillation column simulation, enabling the direct optimization of multiple candidate fluids in PFO, as shown in \cite{Sethi2025}.

Regardless of the optimization strategy, accurate property prediction is paramount.
The literature showcases a wide spectrum of methods, from quantum chemistry-based predictions \cite{Lehmann2004,Struebing2013,Fleitmann2023,Polte2023} to SAFT-based models \cite{Stavrou2014}.
More recently, machine learning has emerged as a dominant tool. There has been a surge of interest in developing artificial neural network (ANN) models for predicting key thermodynamic properties, particularly activity coefficients at infinite dilution, which are crucial for separation process design \cite{Bestwick2023}.
Notable recent examples include models that incorporate physical laws, such as the Gibbs-Duhem equation, to ensure thermodynamic consistency \cite{Winter2023,Rittig2023,DiCaprio2023,Specht2024} and various graph neural network (GNN) architectures \cite{Rittig2023,Medina2023a,Medina2023b}.
This highlights the ongoing drive for faster and more accurate property prediction.

Our work features a demonstration of the developed modelfluid representation on the specific challenge of entrainer selection for enhancing distillation-based separations.
This is a well-established problem in PFO, with recent studies exploring integrated entrainer screening and separation sequence synthesis \cite{Meng2024} and optimal solvent design for extractive distillation \cite{Zhou2019}.
Most studies in PFO rely on commercial process simulators like Aspen \cite{Zhou2019} or in-house tools \cite{Stavrou2023,Loth2023}, demonstrating the practical relevance of the problem.
Also, there is a trade-off found in existing PFO methods: the discrete, group-contribution approaches offer chemical intuitiveness but face combinatorial complexity, while the continuous, SAFT-based methods are powerful but can be less interpretable and may require a pre-restricted search space.
This motivates our work to develop a new modelfluid representation tailored specifically for distillation.
Our goal is to develop a representation based on directly interpretable, physical features of the vapor liquid equilibrium, thereby combining optimization power with clear, cause-and-effect relationships.
Also, we choose modelfluid features that are highly available in databases or can be predicted.

\section{A Modelfluid for Distillation-Based Processes \label{sec:modelfluid}}
This section introduces a novel modelfluid representation tailored for distillation-based processes.
Our framework is built upon physically intuitive \textit{features} which are the primary, optimizable descriptors that define a fluid's behavior.
In the following, we motivate the choice of modelfluid features and explain their direct use in MESH-equation based distillation column simulation, followed by a discussion on the accuracy of the proposed modelfluid representation.

\subsection{Modelfluid Features \label{sec:modelfluidfeatures}}
Modelfluid features are the unique descriptors that identify a fluid mixture and serve as the degrees of freedom in fluid design optimization.
To be effective, these features should be continuous, deterministic, possess physical dimensions or empirical bounds, and be readily available or predictable (see requirements in \sectionref{sec:introduction}).
Critically, they must enable the modeling of key phenomena in distillation: vapor pressure behavior, mixing non-ideality, separation effort, and system enthalpy.
We aim for a minimal set of features that provides a sufficient compromise between model complexity and accuracy for our intended case studies.

For clarity, we will first define the features for a binary mixture and then demonstrate the extension to multi-component systems.
A central concept we use in this work is that of \textit{infinite dilution}.
We adopt the following shorthand notation for some property $a$ of component $i$ at infinite dilution in component $j$:
\begin{equation} \label{eq:infinitedilutionnotation}
    \cdot_i \vert_{j} \equiv \cdot_i\inb{a_j \to 1}
\end{equation}

The selected features are presented below, grouped by the physical property they represent.

\paragraph{Pure Component and System Properties}
The fundamental properties of the pure components and the system are defined by:
\begin{itemize}
    \item The saturated vapor temperatures of the pure components at a given pressure, which intuitively anchor the system's boiling behavior (requirement \textbf{R3}).
        \begin{equation} \label{eq:modelfluidfeatures:saturatedvaportemperature}
            \temperature_i^{\saturatedvapor} = \temperature\vert_i \quad \forall i \in \left[1,N\right]
        \end{equation}
    \item The system pressure $\pressure$, which is essential for ensuring a unique mapping between feature sets and real systems, as explained later in this section in more detail.
    \item The vaporization enthalpy of the pure components at their saturation temperature, which provides a first-order approximation of the energy required for separation.
        \begin{equation} \label{eq:modelfluidfeatures:vaporizationenthalpy}
            \vaporizationenthalpyof{i}\inb{\saturatedvaportemperatureof{i}} \quad \forall i \in \left[1,N\right]
        \end{equation}
\end{itemize}

\paragraph{Mixture Interaction Properties}
The behavior of the components within the mixture is captured by properties at infinite dilution:
\begin{itemize}
    \item The activity coefficient at infinite dilution, which quantifies the non-ideality of the liquid phase.
        \begin{equation} \label{eq:modelfluidfeatures:activitycoefficient}
            \ln\inb{\activitycoefficient_i\vert_j} \quad \forall i, j \in \left[1,N\right], i \neq j
        \end{equation}
        This choice is strategic.
        While mixture data is often scarce, reliable methods exist to predict activity coefficients at infinite dilution from molecular structure (SMILES) or by completing sparse data matrices, as reviewed in \sectionref{sec:literaturereview}, thus satisfying the availability requirement (\textbf{R1}).
        Furthermore, these features directly inform common solvent screening metrics like \textit{capacity} and \textit{selectivity} \cite{Zhou2019}.

    \item The derivative of the vapor mole fraction with respect to the liquid mole fraction at infinite dilution.
        \begin{equation} \label{eq:modelfluidfeatures:vapormolarfractionderivative}
            \frac{\partial \vapormolarfractionof{i}}{\partial \liquidmolarfractionof{i}}\vert_j \quad \forall i, j \in \left[1,N\right], i \neq j
        \end{equation}
        This derivative-based feature is a powerful and direct measure of volatility.
        As shown in \cite{Bortz2015}, it is related to Henry's Law constant, $K_i^H$.
        Applying L'Hôpital's rule to the definition of Henry's constant, we find:
        \begin{equation} \label{eq:henryslawconstant}
            K_i^H = \lim_{\liquidmolarfractionof{i}\to 0} \frac{\vapormolarfractionof{i} \cdot \pressure}{\liquidmolarfractionof{i}}\vert_j = \frac{\partial \vapormolarfractionof{i}}{\partial \liquidmolarfractionof{i}}\vert_j \cdot \pressure
        \end{equation}
        This allows for the calculation of the relative volatility at infinite dilution, $\alpha_{ij}$, an indicator of separation difficulty \cite{Biegler1997, Meng2024}:
        \begin{equation}
            \alpha_{ij} = \frac{K_i^H\vert_j}{K_j^H\vert_i} = \frac{\partial \vapormolarfractionof{i}}{\partial \liquidmolarfractionof{i}}\vert_j \cdot \left(\frac{\partial \vapormolarfractionof{j}}{\partial \liquidmolarfractionof{j}}\vert_i\right)^{-1}
        \end{equation}
        This link to relative volatility further satisfies requirement \textbf{R3} by providing an intuitive measure of distillability.
        For a detailed thermodynamic treatment of Henry's Law, the interested reader is referred to \cite[Ch. 10]{Prausnitz1998}.
\end{itemize}

For a binary system ($N=2$), the complete set of modelfluid features is:
\begin{equation} \label{eq:modelfluidfeatures}
    \modelfluidfeatures = \left[
        \saturatedvaportemperatureof{1},
        \saturatedvaportemperatureof{2},
        \activitycoefficient_1\vert_2,
        \activitycoefficient_2\vert_1,
        \frac{\partial \vapormolarfractionof{1}}{\partial \liquidmolarfractionof{1}}\vert_2,
        \frac{\partial \vapormolarfractionof{2}}{\partial \liquidmolarfractionof{2}}\vert_1,
        \vaporizationenthalpyof{1}\inb{\saturatedvaportemperatureof{1}},
        \vaporizationenthalpyof{2}\inb{\saturatedvaportemperatureof{2}},
        \ln\inb{\pressure}
    \right]
\end{equation}

\paragraph{Extension to Multi-Component Systems}
Extending the feature set to multi-component systems, such as a ternary mixture ($N=3$), is mostly straightforward. We simply define the features for each binary pair within the system: $(1,2)$, $(1,3)$, and $(2,3)$.
For instance, component $1$ will now have activity coefficients at infinite dilution in both component $2$ ($\activitycoefficient_1\vert_2$) and component $3$ ($\activitycoefficient_1\vert_3$).
This leads to a set of 19 features for a ternary system (3 $\temperature_i^{\saturatedvapor}$, 6 $\activitycoefficient_i\vert_j$, 6 $\frac{\partial \vapormolarfractionof{i}}{\partial \liquidmolarfractionof{i}}\vert_j$, 3 $\vaporizationenthalpyof{i}$, and 1 $\ln\inb{\pressure}$).
However, due to underlying thermodynamic consistency, not all these features are independent.
This will be discussed at the end of \sectionref{sec:modelfluidparameters}.

\subsection{From Modelfluid Feature to Process Simulation \label{sec:modelfluidparameters}}
While the features provide an intuitive and optimizable description of a fluid, they cannot be used directly in standard process simulators.
To bridge this gap and satisfy requirement \textbf{R2}, we derive the usage of the modelfluid features in established thermodynamic models.
This mapping makes our modelfluid representation a practical tool for flowsheet simulation, as demonstrated in \appendixref{sec:fakeenthalpysimulation}.

To perform the vapor liquid equilibria- and enthalpy computations necessary in distillation column simulation, we rely on three core models and the system's state.

\paragraph{1. Activity Coefficient Model} We use the Three-Suffix Margules model \cite{Prausnitz1998}
\begin{equation}
    \label{eq:margules}
    \ln\inb{\activitycoefficient_i\inb{\liquidmolarfractions}} = \inb{\ln\inb{\activitycoefficient_i}\vert_{j} + 2 \cdot \inb{\ln\inb{\activitycoefficient_j}\vert_{i} - \ln\inb{\activitycoefficient_i}\vert_{j} \cdot \liquidmolarfractionof{i}}} \cdot \liquidmolarfractionof{j}^2 ,
\end{equation}
which is fully parameterized by the activity coefficients at infinite dilution, which are part of the modelfluid features \eqref{eq:modelfluidfeatures:activitycoefficient}.
This synergy is by design, enabling the connection of our framework with modern prediction methods for $\activitycoefficient_i\rvert_j$ \cite{Jirasek2020} -- in the situation where experimental data or existing models are not available.
Due to the composition-dependent activity coefficient modeling by the Margules model, no constant relative volatility needs to be used in the distillation column simulation, see \appendixref{sec:fakeenthalpysimulation}.
We acknowledge that the Margules model assumes temperature-independent activity coefficients, which is an acceptable simplification for many distillation systems, in particular such obeying azeotropic phase behavior.

\paragraph{2. Vapor Pressure Model} We model the vapor pressure of each pure component using a simplified two-parameter variant of the Antoine equation \cite{Wilding1998}:
\begin{equation} \label{eq:simplifiedantoine}
    \ln\inb{\propertyindex{\propertyphase{\pressure}{\saturatedvapor}}{i}\inb{\temperature}} = \parameter^{\simplifiedantoine}_{i,1} + \frac{\parameter^{\simplifiedantoine}_{i,2}}{\temperature} .
\end{equation}
It's parameters, $\parameter^{\simplifiedantoine}_{i,1}$ and $\parameter^{\simplifiedantoine}_{i,2}$, are obtained using an explicit map from the modelfluid features:

Since, at the saturated vapor temperature (which is part of the modelfluid features), the vapor pressure of a pure fluid substance is equal to the pressure of the system,
we obtain the first Antoine equation parameter as follows:
\begin{equation} \label{eq:modelfluidmapping:linearsystem:vaporpressure}
    \parameter_i^{\simplifiedantoine, 1} = \ln\inb{\pressure} - \frac{\parameter_i^{\simplifiedantoine, 2}}{\temperature_i^{\saturatedvapor}} \quad \forall i \in \{1, 2\}.
\end{equation}
Then, we use the logarithmic variant of extended Raoult's Law \eqref{eq:extendedraoultslaw}
\begin{equation} \label{eq:extendedraoultslawlogarithmic}
    \ln\inb{\vapormolarfractionof{i}} + \ln\inb{\pressure} = \ln\inb{\liquidmolarfractionof{i}} + \ln\inb{\activitycoefficient_i\inb{\liquidmolarfractions, \temperature}} + \ln\inb{\pressure_i^{\saturatedvapor}\inb{\temperature}} ,
\end{equation}
for which we consider its derivative w.r.t. the liquid-phase molar fractions at the point of infinite dilution in component $j$:
\begin{equation} \label{eq:modelfluidmapping:linearsystem:derivative}
    \ln\inb{\frac{\partial \vapormolarfractionof{i}}{\partial \liquidmolarfractionof{i}}\bigg\vert_j} + \ln\inb{\pressure} = \parameter_i^{\simplifiedantoine, 1} + \frac{\parameter_i^{\simplifiedantoine, 2}}{\temperature_j^{\saturatedvapor}} + \ln\inb{\activitycoefficient_i}\vert_{j} \quad \forall i, j \in \{1, 2\}, i \neq j.
\end{equation}
Antoine equation parameter $\parameter_i^{\simplifiedantoine, 2}$ is obtained by rearranging \eqref{eq:modelfluidmapping:linearsystem:derivative}.

\paragraph{3. Enthalpy Model} The enthalpy of vaporization is approximated as a composition-weighted average of the pure component features, assuming no heat of mixing and temperature-independent enthalpies:
\begin{equation} \label{eq:vaporizationenthalpymodel}
    \molarenthalpy_{mix}^{\vaporization} = \sum_{i=1}^{N} \liquidmolarfractionof{i} \cdot \vaporizationenthalpyof{i}\inb{\saturatedvaportemperatureof{i}}
\end{equation}
The pure component vaporization enthalpies $\vaporizationenthalpyof{i}\inb{\saturatedvaportemperatureof{i}}$ are part of modelfluid features.

\paragraph{Pressure} The pressure $\pressure$ is an essential feature of the modelfluid as it is maintaining the identifiability of the system.
For example, for some system A+B at pressure $\pressure_1$ all modelfluid features $\modelfluidfeatures$ except for the pressure could -- in theory -- be identical to those of system A+C at pressure $\pressure_2$, for $B \neq C$.\\
%
%

Ultimately, the assumptions made here represent a deliberate trade-off between model fidelity and feature space dimensionality.
The modelfluid framework is flexible and can be extended with more complex models or additional features should an application require greater accuracy.

\paragraph{Extension to Multi-Component Systems} The extension to multi-component systems is generally analogous to the extension of the features described in \sectionref{sec:modelfluidfeatures}.
For a ternary system, we employ the ternary Margules model \cite[p. 283]{Prausnitz1998}.
A critical consideration, detailed in \appendixref{sec:modelfluidparameters:reduction}, is the enforcement of physical consistency for pure component properties.
The vapor pressure model for a given component must be unique regardless of the binary pair it is in -- a condition that sounds trivial but is easily overlooked.
This constraint reduces the number of independent features from $19$ to $16$ for a ternary system, as shown in \appendixref{sec:modelfluidparameters:reduction}.

\subsection{Modelfluid Error} \label{sec:modelfluiderror}
As established in the preceding sections, the proposed modelfluid representation is founded on physically interpretable features derived from the vapor liquid equilibrium.
These features are mapped via an explicit mapping to the parameters of established empirical models (a vapor pressure model and an activity coefficient model), ensuring integrability into existing simulation tools.
While the physical motivation and usability of this representation have been discussed, a quantitative assessment of its accuracy is essential.

The error analysis naturally focuses on the VLE, as it is the basis for the modelfluid features.
However, a direct analytical treatment of the error is non-trivial.
The process involves transformations: from some reference VLE - computed using thermodynamic models, e.g. the NRTL activity coefficient model - to the modelfluid features ($\feature^{\modelfluid}$), to the re-computed VLE based on the parameterization of the Margules activity coefficient model and the simplified Antoine equation for vapor pressure, as described in \sectionref{sec:modelfluidparameters}.
The implicit nature of the governing equation, extended Raoult's Law \eqref{eq:extendedraoultslaw}, further complicates a direct analysis.

To overcome this, we derive an approximate, explicit expression for the error in the VLE bubble point temperature, denoted as $\delta \temperature$.
For the sake of brevity, we only present the final expression here but provide a detailed derivation in \appendixref{sec:modelingerror}.

To simplify notation, we use $\modelfluidparameters=\parameter\inb{\modelfluidfeatures}$ to denote the thermodynamic modeling as presented in \sectionref{sec:modelfluidparameters}, and $\parameter^{\rigorous}$ to denote an alternative reference modeling which is used to determine the modelfluid features and to which we consider the modeling error.

Furthermore, to arrive at a tractable expression, we make a simplifying assumption: we consider the rigorous and modelfluid vapor pressure models to be equivalent.
This is a reasonable approximation for the purpose of error analysis, as the simplified Antoine model \eqref{eq:simplifiedantoine} captures the dominant temperature dependency of a component's vapor pressure.
We denote this unified vapor pressure model as $\bar{\pressure}_i^{\saturatedvapor}\inb{\temperature}$.

Under this assumption, and using the relation from \eqref{eq:partialtemperaturet} from \appendixref{sec:modelingerror}, the temperature error can be expressed as:
\begin{equation} \label{eq:modelfluid:error:temperature}
    \delta \temperature\inb{\liquidmolarfractions, \temperature, \parameter^{\rigorous}} = \frac{\sum_{i=1}^{\mynum_{\component}} \liquidmolarfractionof{i} \cdot \bar{\pressure}_i^{\saturatedvapor}\inb{\temperature} \cdot \inb{\activitycoefficient_i^{\modelfluid}\inb{\liquidmolarfractions, \parameter^{\modelfluid}} - \activitycoefficient_i^{\rigorous}\inb{\liquidmolarfractions, \temperature, \parameter^{\rigorous}}}}{\sum_{i=1}^{\mynum_{\component}} \liquidmolarfractionof{i} \cdot \bar{\pressure}_i^{\saturatedvapor}\inb{\temperature} \cdot \inb{-\frac{\parameter_i^{\simplifiedantoine, 2}}{\temperature^2}} \cdot \activitycoefficient_i^{\modelfluid}\inb{\liquidmolarfractions, \parameter^{\modelfluid}}} .
\end{equation}

We can now analyze the magnitude of this error.
The numerator is driven by the difference between the modelfluid (Margules) and the rigorous activity coefficient models, $\activitycoefficient_i^{\modelfluid} - \activitycoefficient_i^{\rigorous}$.
By construction, our method ensures this difference is zero at the pure component limits (i.e., as $\liquidmolarfractionof{i} \to 0$ and $\liquidmolarfractionof{i} \to 1$) and is designed to be small across the intermediate composition range.
The denominator, on the other hand, contains the Antoine parameter-temperature-dependent term $-\frac{\parameter_i^{\simplifiedantoine, 2}}{\temperature^2}$.
From the definition in \eqref{eq:simplifiedantoine}, the parameter $\parameter_i^{\simplifiedantoine, 2}$ is negative ($\parameter_i^{\simplifiedantoine, 2} < 0$) for any real fluid, as vapor pressure increases with temperature.
Furthermore, its magnitude is typically large, reflecting the strong dependency of vapor pressure on temperature.

Consequently, the structure of \eqref{eq:modelfluid:error:temperature} - a small numerator divided by a denominator with a large-magnitude term - suggests that the resulting temperature error $\delta \temperature$ will be small.
This conclusion is corroborated by the numerical examples presented in \appendixref{sec:modelingerror}, where we observe only minor deviations between the rigorous and modelfluid VLE calculations for a vast majority of systems considered in this study.

Note that while we only discuss the temperature error in this section, the error in the vapor phase composition $\vapormolarfractionof{i}$ is straightforward to compute using \eqref{eq:extendedraoultslaw} and the expression for the temperature error $\delta \temperature$ \eqref{eq:modelfluid:error:temperature}.

\section{Optimal Entrainer Search \label{sec:entrainersearch}}
\subsection{Case study: Entrainer Distillation} \label{sec:entrainerdistillation}

Entrainer distillation is a specialized separation process for binary liquid mixtures that form azeotropes.
At the azeotropic point, the liquid and vapor phases have identical compositions, which prevents separation by simple distillation.
The process introduces a third component, the \textit{entrainer}, to break the azeotrope by altering the phase behavior, thus enabling the separation. 
The process flowsheet used in this work is depicted in Fig.~\ref{fig:entrainer-distillation-flowsheet}.
Depending on the azeotrope system (minimum- or maximum-boiling), different column sequences and recycle streams are possible.
For a comprehensive overview of entrainer distillation, see \cite{Duessel1995}.
\usetikzlibrary{arrows.meta, positioning, calc}

\newcommand{\productStreamXLength}{1.5}
\newcommand{\productStreamYLength}{0.5}
\newcommand{\columnXLength}{1.5}
\newcommand{\columnYLength}{4}
\newcommand{\unitDistance}{2.5}
\newcommand{\mixerLength}{1}
\newcommand{\streamspec}[1]{$\liquidflow_{#1},\liquidmolarfractions_{#1}$}

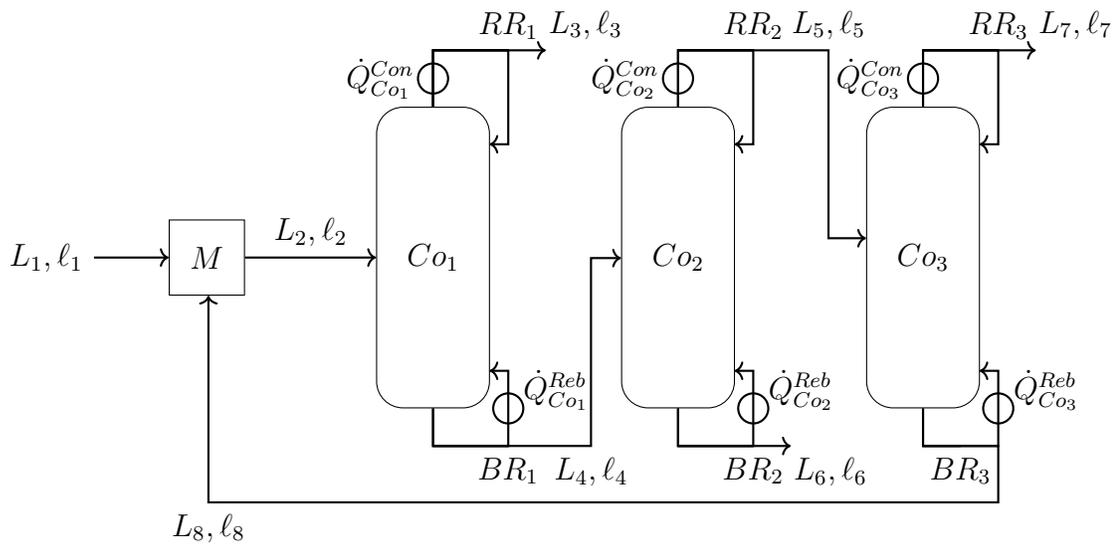
\begin{figure}
    \centering
        
    \begin{tikzpicture}[auto]

        \tikzstyle{column} = [draw, rectangle, minimum width=\columnXLength cm, minimum height=\columnYLength cm, rounded corners=10pt]
        \tikzstyle{mixer} = [draw, rectangle, minimum size=\mixerLength cm]
        \tikzstyle{stream} = [->, thick]

        \node[mixer] (M) at (0,0) {$M$};

        \node[column] (C1) at ($(M.east) + (\unitDistance,0)$) {$\column_1$};
        \node[column] (C2) at ($(C1.east) + (\unitDistance,0)$) {$\column_2$};
        \node[column] (C3) at ($(C2.east) + (\unitDistance,0)$) {$\column_3$};

        \draw[stream] (-\productStreamXLength,0) node[left] {\streamspec{1}} -- (M);
        \draw[stream] (M) -- node[above] {\streamspec{2}} (C1);

        \draw[stream] (C1.north) -- ++(0,\productStreamYLength*1.5) -- ++(\productStreamXLength,0) node[above, xshift=\productStreamXLength*0.35 cm] {\streamspec{3}};
        \draw[stream] (C1.south) -- ++(0,-\productStreamYLength) -- ++(\unitDistance/2+\productStreamXLength/2 + 0.1,0) node[below] {\streamspec{4}} -- ++(0,\productStreamYLength+\columnYLength/2) --++ (\unitDistance/2-\productStreamXLength/2 - 0.1,0);

        \draw[stream] (C1.north) -- ++(0,\productStreamYLength*1.5) -- ++(\productStreamXLength*2/3,0) -- ++(0,-\productStreamYLength*1.5) -- ++(0,-\productStreamYLength) -- ++(-\productStreamXLength*2/3+\columnXLength/2,0);
        \draw[thick] (C1.north) -- ++(0,\productStreamYLength*1.5) -- ++(\productStreamXLength*2/3,0) node[above] {$\refluxratio_1$};
        \draw[stream] (C1.south) ++(0,-\productStreamYLength) -- ++(\productStreamXLength*2/3,0) -- ++(0,\productStreamYLength) node[right, xshift=\productStreamXLength, yshift=0.25 cm] {$\reboilerduty_{\column_1}$} -- ++(0,\productStreamYLength) -- ++(-\productStreamXLength*2/3+\columnXLength/2,0);
        \draw[thick] (C1.south) -- ++(0,-\productStreamYLength) -- ++(\productStreamXLength*2/3,0) node[below] {$\boilupratio_1$};

        \draw[thick] (C1.north) ++(0,\productStreamYLength*0.75) circle (0.2);
        \draw[thick] (C1.north) ++(0,\productStreamYLength*0.75) node[left, xshift=-0.1 cm] {$\condenserduty_{\column_1}$};
        \draw[thick] (C1.south) ++(0,-\productStreamYLength) ++(\productStreamXLength*2/3,0) ++(0,\productStreamYLength) circle (0.2);

        \draw[stream] (C2.north) -- ++(0,\productStreamYLength*1.5) -- ++(\unitDistance/2+\productStreamXLength/2,0) node[above] {\streamspec{5}} -- ++(0,-\productStreamYLength-\columnYLength/2) --++ (\unitDistance/2-\productStreamXLength/2,0);
        \draw[stream] (C2.south) -- ++(0,-\productStreamYLength) -- ++(\productStreamXLength,0) node[below, xshift=\productStreamXLength*0.35 cm] {\streamspec{6}};

        \draw[stream] (C2.north) ++(0,\productStreamYLength*1.5) -- ++(\productStreamXLength*2/3,0) -- ++(0,-\productStreamYLength*1.5) -- ++(0,-\productStreamYLength) -- ++(-\productStreamXLength*2/3+\columnXLength/2,0);
        \draw[thick] (C2.north) -- ++(0,\productStreamYLength*1.5) -- ++(\productStreamXLength*2/3,0) node[above] {$\refluxratio_2$};
        \draw[stream] (C2.south) ++(0,-\productStreamYLength) -- ++(\productStreamXLength*2/3,0) -- ++(0,\productStreamYLength) node[right, xshift=\productStreamXLength, yshift=0.25 cm] {$\reboilerduty_{\column_2}$} -- ++(0,\productStreamYLength) -- ++(-\productStreamXLength*2/3+\columnXLength/2,0);
        \draw[thick] (C2.south) -- ++(0,-\productStreamYLength) -- ++(\productStreamXLength*2/3,0) node[below] {$\boilupratio_2$};

        \draw[thick] (C2.north) ++(0,\productStreamYLength*0.75) circle (0.2);
        \draw[thick] (C2.north) ++(0,\productStreamYLength*0.75) node[left, xshift=-0.1 cm] {$\condenserduty_{\column_2}$};
        \draw[thick] (C2.south) ++(0,-\productStreamYLength) ++(\productStreamXLength*2/3,0) ++(0,\productStreamYLength) circle (0.2);

        \draw[stream] (C3.north) -- ++(0,\productStreamYLength*1.5) -- ++(\productStreamXLength,0) node[above, xshift=\productStreamXLength*0.35 cm] {\streamspec{7}};
        \draw[stream] (C3.south) -- ++(0,-\productStreamYLength) -- ++(\productStreamXLength*2/3,0) -- ++(0,-\productStreamYLength*1.5) -- ++(-\productStreamXLength*2/3-\unitDistance*3-\columnXLength-\mixerLength/2,0) node[below] {\streamspec{8}} -- ++(0,\productStreamYLength*2.5+\columnYLength/2-\mixerLength/2);

        \draw[stream] (C3.north) ++(0,\productStreamYLength*1.5) -- ++(\productStreamXLength*2/3,0) -- ++(0,-\productStreamYLength*1.5) -- ++(0,-\productStreamYLength) -- ++(-\productStreamXLength*2/3+\columnXLength/2,0);
        \draw[thick] (C3.north) -- ++(0,\productStreamYLength*1.5) -- ++(\productStreamXLength*2/3,0) node[above] {$\refluxratio_3$};
        \draw[stream] (C3.south) ++(0,-\productStreamYLength) -- ++(\productStreamXLength*2/3,0) -- ++(0,\productStreamYLength) node[right, xshift=\productStreamXLength, yshift=0.25 cm] {$\reboilerduty_{\column_3}$} -- ++(0,\productStreamYLength) -- ++(-\productStreamXLength*2/3+\columnXLength/2,0);
        \draw[thick] (C3.south) -- ++(0,-\productStreamYLength) -- ++(\productStreamXLength*1/3,0) node[below] {$\boilupratio_3$};

        \draw[thick] (C3.north) ++(0,\productStreamYLength*0.75) circle (0.2);
        \draw[thick] (C3.north) ++(0,\productStreamYLength*0.75) node[left, xshift=-0.1 cm] {$\condenserduty_{\column_3}$};
        \draw[thick] (C3.south) ++(0,-\productStreamYLength) ++(\productStreamXLength*2/3,0) ++(0,\productStreamYLength) circle (0.2);

    \end{tikzpicture}
    
    \caption{
        Entrainer distillation flowsheet for the separation of a binary mixture with maximum-boiling azeotropic phase behavior.
        The columns are denoted by $\column_1$, $\column_2$, and $\column_3$, and the mixer is denoted by $M$.
        The liquid flow rates and molar fractions of the streams are represented by $\liquidflow$ and $\liquidmolarfractions$, respectively.
        The subscripts of these variables correspond to the stream numbers shown in the figure.
        The variables $\refluxratio$, $\boilupratio$, $\condenserduty$, $\reboilerduty$, and $\heatduty$ represent the reflux ratio, boilup ratio, condenser duty, reboiler duty, and heat duty, respectively.
    }
    \label{fig:entrainer-distillation-flowsheet}

\end{figure}



We chose entrainer distillation as a case study for several key reasons.
It presents a challenging, industrially relevant optimization problem that is well-known in chemical process engineering.
Selecting the best entrainer, though, is not a simple screening task but a coupled process-and-fluid optimization where operating conditions must be re-optimized for each candidate.
This complexity provides an ideal test case for our modelfluid representation.
To discuss the optimality of different entrainer candidates and operating points, we use a classic engineering trade-off between capital expenditure (CAPEX), represented by the total number of stages ($\mynum_{\stages}^{\total}$), and operational expenditure (OPEX), represented by the total heat duty ($\totalreboilerduty$) -- also known as the NQ curve \cite{Biegler1997}.

\paragraph{Scope and Assumptions}

Our primary goal is to demonstrate a novel modelfluid representation on the example of entrainer selection, not to perform an exhaustive optimization of the distillation process itself. 
Therefore, we adopt the following simplifications and assumptions:

\begin{itemize}
    \item \textbf{Phase Behavior:} We only consider entrainers that form homogeneous, stable liquid phases. Systems that exhibit vapor-liquid-liquid equilibrium (VLLE) or heterogeneous azeotropes are excluded, which is a limitation of our current simulation framework.
    
    \item \textbf{Process Structure:} We do not perform superstructure optimization or consider special two-column separation sequences.
    
    \item \textbf{Simulation Model:}
    To be able to use the thermodynamic model of the modelfluid representation as sketched in \sectionref{sec:modelfluid}, we implemented our own process simulation.
    It is based on the modeling described in \cite{Hoffmann2017}, but features modelfluid-tailored adaptations which are described in \appendixref{sec:fakeenthalpysimulation}.
    Therefore, all optimization results presented in this study are obtained using the simulation framework described in \appendixref{sec:fakeenthalpysimulation}.
    This includes mapping the modelfluid features \eqref{eq:modelfluidfeatures} to the parameters of the two-parameter Antoine vapor pressure equation and the Margules activity coefficient model, as described in \sectionref{sec:modelfluid}.
    Note that the above approach does not use constant relative volatilities.
    \item \textbf{Operating Conditions and Thermodynamics:}
    \begin{enumerate}
        \item Streams outside columns are at their bubble point.
        \item All columns operate at the same constant, uniform pressure (no pressure drop over stages).
        \item Columns are adiabatic, with heat exchange only at the condenser and reboiler.
        \item Total condensation (evaporation) is assumed for the condensers (reboilers).
        \item The enthalpy of mixing is neglected.
    \end{enumerate}
\end{itemize}

\subsection{Finding the Best Entrainer Candidates} \label{sec:entrainerhypothetical}

The conventional method for selecting a suitable entrainer relies on extensive screening studies, a combinatorial task that is often prohibitively expensive.
Our approach circumvents this by employing a continuous optimization strategy to identify a \textit{hypothetically optimal entrainer}.
This is enabled by our modelfluid representation, which allows the optimizer to continuously navigate the space of entrainer features.
The goal is to find an optimal set of modelfluid features which, while perhaps not corresponding to a real molecule, serves as an ideal target to guide the selection of the best available candidate from a given pool.

\begin{figure*}[t!]

\centering
\begin{adjustbox}{width=\textwidth, center, scale=0.7}
\begin{tikzpicture}[
    node distance=10mm and 15mm,
    process/.style={
        rectangle, 
        rounded corners=3mm, 
        draw, 
        thick, 
        fill=blue!10, 
        align=center,
        minimum height=15mm, 
        minimum width=30mm,
        drop shadow
    },
    data/.style={
        trapezium, 
        trapezium left angle=70, 
        trapezium right angle=110,
        draw, 
        thick, 
        fill=green!15, 
        align=center, 
        minimum height=10mm
    },
    database/.style={
        cylinder, 
        shape border rotate=90, 
        aspect=0.25, 
        draw, 
        thick, 
        fill=orange!20,
        align=center, 
        minimum height=15mm, 
        minimum width=25mm
    },
    digitaltwin/.style={
        rectangle, 
        rounded corners=3mm, 
        draw=red!80, 
        thick, 
        fill=red!10, 
        align=center,
        minimum height=20mm, 
        minimum width=40mm,
        drop shadow
    },
    arrow/.style={
            -Stealth,
            thick
        },
    label/.style={
        text=blue!80!black, 
        font=\small\bfseries
    },
    fitbox/.style={
        rectangle,
        rounded corners=5mm,
        draw=gray!50,
        thick,
        dashed
    }
]

\node (reference_nq_curves) [data] {\textbf{Step 1}: \\ Compute baseline \\ NQ curve of \\ the reference entrainer};
\node (hypofluid) [digitaltwin, right=of reference_nq_curves] {Hypothetical \\ optimal entrainer};
\node (pfo) [process, above=of hypofluid, yshift=5mm] {Add entrainer features to \\ the optimization variables};
\draw[arrow] (reference_nq_curves) -- (pfo) node[midway, above, font=\small, fill=white, yshift=-3mm] {Fix column sizes};
\draw[arrow] (pfo) -- (hypofluid) node[midway, above, font=\small, fill=white, yshift=-3mm] {Optimize};

\node (step2_box) [fitbox, fit=(pfo) (hypofluid), inner sep=8mm] {};
\node[label, above=1mm of step2_box.north] {Step 2: Process Fluid Optimization};

\node (postoptim) [process, below=of hypofluid, yshift=-5mm] {Compute the \\ hypothetical NQ curve};
\draw[arrow] (hypofluid) -- (postoptim);

\node (foreach_box) [fitbox, fit=(pfo) (hypofluid) (postoptim), inner sep=8mm] {};
\node[label, below=1mm of foreach_box.south] {For each point on the NQ curve};

\draw[arrow] (postoptim) -- (reference_nq_curves) node[midway, above, font=\small, fill=white, yshift=-3mm] {Compare in \figureref{fig:NQValidation:Hypo}};

\end{tikzpicture}
\end{adjustbox}
\caption{
    Schematic workflow for the identification of hypothetical optimal entrainer components.
    Starting from a known reference entrainer's NQ curve, we compute the optimal entrainer features for each point on the NQ curve, fixing the number of stages in the column and further minimizing the heat duty in the flowsheet.
    This yields a set of hypothetical optimal entrainer features, each representing an individual hypothetical entrainer.
}
    \label{fig:entraineroptimization}
\end{figure*}
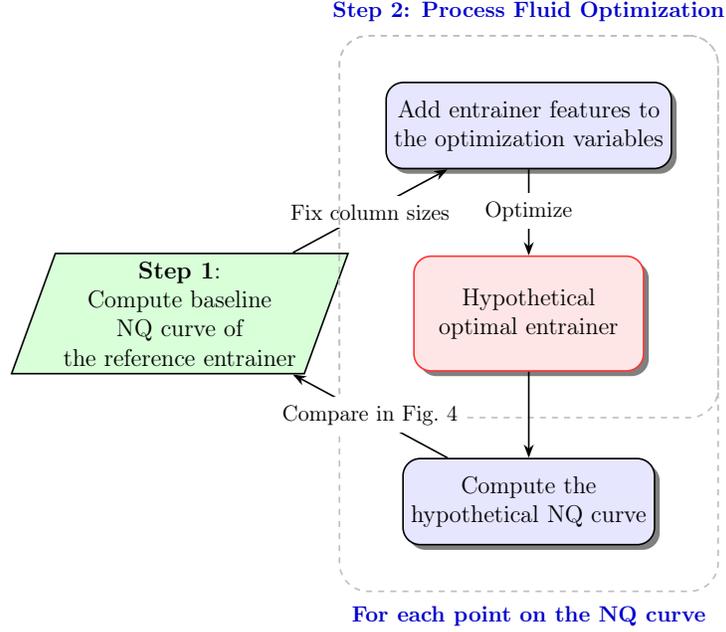

As sketched in \figureref{fig:entraineroptimization}, our methodology unfolds in two main steps.
First, we establish a performance baseline for the separation process using a known reference entrainer.
Second, using this baseline, we optimize the entrainer's modelfluid features to find the properties of a hypothetically optimal entrainer.
Once one or more hypothetical optimal entrainers are identified, we can screen a database of real molecules to find the best available candidate that closely matches the hypothetical features - which we discuss in \sectionref{sec:entraineravailable}.

For our case study, we consider the separation of the maximum-boiling azeotrope of \textit{Acetone} and \textit{Chloroform}.
We use \textit{Benzene} as the reference entrainer, against which we will benchmark the performance of our optimized hypothetical candidates.

\subsubsection{Step 1: Baseline Process Optimization (NQ Curve)}

To establish a baseline, we first perform a multi-objective optimization of the entrainer distillation flowsheet using the reference entrainer (Benzene).
The goal is to find the Pareto-optimal trade-off between the total number of theoretical stages ($\mynum_{\stages}^{\total}$), which is a proxy for investment costs, and the total reboiler duty ($\totalreboilerduty$), a proxy for operating costs.
The resulting Pareto frontier is often referred to as an NQ curve.

\paragraph{Optimization Formulation.}
The problem is formulated to simultaneously minimize the two competing objectives:
\begin{equation} \label{eq:entrainerdistillation:objectivefunction:totalnumberofstages}
    \mynum_{\stages}^{\total} = \sum_{i=1}^{\mynum_{\column}} (\mynum_{\stages, \column_i}^{\belowfeed} + 1 + \mynum_{\stages, \column_i}^{\abovefeed})
\end{equation}
\begin{equation} \label{eq:entrainerdistillation:objectivefunction:totalheatduty}
    \totalreboilerduty = \sum_{i=1}^{\mynum_{\column}} \reboilerduty_{\column_i}
\end{equation}
These objectives are minimized by varying a set of process variables, $\feature_{\process}$, defined as:
\begin{align} \label{eq:entrainerdistillation:optimizationvariables}
    \feature_{\process} = 
    \left[
    \begin{aligned}
        &\spl_{\column_i}, \refluxratio_{\column_i}, \mynum_{\stages, \column_i}^{\belowfeed}, \mynum_{\stages, \column_i}^{\abovefeed}, \\
        &\liquidmolarfraction_{\recycle\component_j\column_i}, \liquidmolarfraction_{\product\component_j\column_i}
    \end{aligned}
    \right]_{i=1.. \mynum_{\column}, j=1..\mynum_{\component}}
\end{align}
This vector includes split fractions, reflux ratios, the number of stages above and below the feed, and the compositions of all product and recycle streams for each of the three columns.
It is noted that the specification of both recycle- $\liquidmolarfraction_{\recycle\component_j\column_i}$ and component $\component_j$ concentrations $\liquidmolarfraction_{\product\component_j\column_i}$ for each column $\column_i$ may be redundant, but is chosen here for numerical reasons, in essence, to avoid the occurrence of negative concentrations during optimization.

This optimization is subject to a set of rigorous constraints to ensure a physically meaningful and feasible process. The product constraints require minimum purities \eqref{eq:entrainerdistillation:productpurity} and molar flow rates \eqref{eq:entrainerdistillation:productflow} for the product component ($\product\component$) stream of each column.
\begin{align} \label{eq:entrainerdistillation:productpurity}
    \liquidmolarfraction_{\product\component\column_i} \geq& \liquidmolarfraction_{\product\component\column_i}^{min} \quad &\forall i = 1, \dots, \mynum_{\column} \\
    \label{eq:entrainerdistillation:productflow}
    \liquidflow_{\product\column_i} \geq& \liquidflow_{\product\column_i}^{min} \quad &\forall i = 1, \dots, \mynum_{\column}
\end{align}
Furthermore, flowsheet mass balances must be satisfied. These include the closure condition that molar fractions in any stream must sum to one \eqref{eq:entrainerdistillation:closureconditions}, the overall component mass balance for the entire flowsheet \eqref{eq:entrainerdistillation:flowsheetmassbalance}, and the component mass balances for each individual column \eqref{eq:entrainerdistillation:columnmassbalance}.
\begin{align} \label{eq:entrainerdistillation:closureconditions}
    1 - \sum_{j=1}^{\mynum_{\component}} \liquidmolarfraction_{\product\component_j\column_i} =& 0 & \forall i = 1, \dots, \mynum_{\column} \\ \nonumber
    1 - \sum_{j=1}^{\mynum_{\component}} \liquidmolarfraction_{\recycle\component_j\column_i} =& 0 & \forall i = 1, \dots, \mynum_{\column}
\end{align}
\begin{equation}  \label{eq:entrainerdistillation:flowsheetmassbalance}
    \liquidflow_1 \cdot \liquidmolarfraction_{1, \component_j} - \sum_{i=1}^{\mynum_{\component}} \liquidflow_{\product\column_i} \cdot \liquidmolarfraction_{\product\component_j\column_i} = 0 \quad \forall j = 1, \dots, \mynum_{\component}
\end{equation}
\begin{align} \label{eq:entrainerdistillation:columnmassbalance}
    &\liquidflow_{\feed\column_i} \cdot \liquidmolarfraction_{\feed\component_j\column_i} - \liquidflow_{\product\column_i} \cdot \liquidmolarfraction_{\product\component_j\column_i} - \liquidflow_{\recycle\column_i} \cdot \liquidmolarfraction_{\recycle\component_j\column_i} = 0 \\ \nonumber
    &\forall i = 1, \dots, \mynum_{\column}, j = 1, \dots, \mynum_{\component}
\end{align}
Finally, the MESH equations governing the stage-to-stage thermodynamic equilibrium within each distillation column, denoted by $\constraint_{\column}$, must be satisfied \eqref{eq:entrainerdistillation:meshconstraint}.
\begin{equation} \label{eq:entrainerdistillation:meshconstraint}
    \constraint_{\column}\inb{\liquidmolarfractions^{{\bottom}}, \spl, \refluxratio, \liquidmolarfractions^{{\feed}}, \liquidflow^{{\feed}}, \pressure, \mynum_{\stages}^{\belowfeed}, \mynum_{\stages}^{\abovefeed}} = 0
\end{equation}
Note that the reboiler duty $\reboilerduty_{\column_i}$ for each column $\column_i$ is computed as part of the MESH equations, using the modelfluid-tailored variant of the stage-to-stage computation as described in \appendixref{sec:fakeenthalpysimulation}.
To generate the Pareto frontier, we solve a series of single-objective problems using the epsilon-constraint method, as the frontier is expected to be non-convex \cite{Nowak2023}. For a given maximum number of stages $\epsilon_{\mynum_{\stages}^{\total}}$, the scalarized problem is formulated as:
\begin{align} \label{eq:entraineroptimizationproblem}
    \feature_{\process}^{\best} =& \arg \min_{\feature_{\process} \in \featurespace_{\process}} \totalreboilerduty \quad \text{s.t.} \quad \mynum_{\stages}^{\total} \leq \epsilon_{\mynum_{\stages}^{\total}} \quad \text{and all other constraints.}
\end{align}
The resulting set of Pareto-optimal solutions provides the starting points for the subsequent entrainer optimization.

\textit{(For brevity, the detailed bounds on the optimization variables are omitted here but listed in \appendixref{app:optimizationdetails}, Step 1.)}

\subsubsection{Step 2: Hypothetical Entrainer Optimization}

Building on this foundation, we now shift our focus to optimizing the entrainer itself.
We consider a scenario where an existing process must be improved by finding a more efficient entrainer.
Therefore, we fix the process structure (i.e., the number of stages in each column) to a specific optimal design point from the NQ curve generated in Step 1.

The optimization problem is then reformulated to find the ideal entrainer features that minimize the total energy consumption for this fixed process design.

We use the Pareto optimal solutions from Step 1 as starting points for this entrainer optimization.
This means that we perform the optimization of entrainer features, as explained in the following, for each of the $n$ points on the NQ curve, yielding between $1$ and $n$ hypothetical optimal entrainers, each corresponding to a specific point on the NQ curve.

\paragraph{Optimization Formulation.}
The optimization variables now include the process operating variables (splits, reflux ratios, stream compositions) and, crucially, the modelfluid features that describe the entrainer.
The set of entrainer-specific modelfluid features, $\feature_{\entrainer}$, includes its pure component properties and its interactions with the solutes:
\begin{equation} \label{eq:entraineroptim:entrainervariables}
    \feature_{\entrainer} = \left[\saturatedvaportemperatureof{2}, \vaporizationenthalpyof{2}, \activitycoefficient_1\vert_2, \activitycoefficient_2\vert_1, \activitycoefficient_2\vert_3, \activitycoefficient_3\vert_2, \frac{\partial \vapormolarfractionof{1}}{\partial \liquidmolarfractionof{1}}\vert_2\right]
\end{equation}
The full set of optimization variables for this problem, $\feature_{\entrainer,\process}$, is formed by combining the entrainer features $\feature_{\entrainer}$ from \eqref{eq:entraineroptim:entrainervariables} with the process variables from \eqref{eq:entrainerdistillation:optimizationvariables}, with the exception that the stage numbers are now fixed parameters.

With the number of stages fixed, the problem becomes a single-objective optimization: to minimize the total heat duty.
\begin{equation}  \label{eq:entraineroptim:objectivefunction}
    \feature_{\entrainer,\process}^{\best} = \arg \min_{\feature_{\entrainer,\process} \in \featurespace_{\entrainer,\process}} \totalreboilerduty .
\end{equation}
This optimization is subject to the same process constraints as in step 1, which includes the MESH equations using the thermodynamics as described in \sectionref{sec:modelfluidparameters}. 
However, to ensure the resulting hypothetical entrainer is physically realistic and to prevent the optimizer from exploiting model artifacts, we introduce additional constraints on the modelfluid features, which are discussed in more detail in \appendixref{sec:thermodynamicfeasibility}:
\begin{itemize}
    \item \textbf{Phase Stability:} To ensure a single homogeneous liquid phase, which is a prerequisite for our simulator, we enforce phase stability conditions.
    Enforcing this for all compositions would yield a semi-infinite optimization problem \cite{Mitsos2015}.
    We approximate this by discretizing the constraint across a grid of compositions, which is sufficient to guide the optimizer away from unrealistic regions.
    See \appendixref{sec:phasestabilityderivation} for a detailed derivation of the phase stability constraints as used in this work.
    \item \textbf{Sub-critical State:} We add constraints to ensure the system operates in a sub-critical state, for example by ensuring the entrainer's boiling point is below the critical temperatures of the solutes. The approximation of the critical temperature is described in \appendixref{sec:thermodynamicfeasibility}.
\end{itemize}

Notably, we do not explicitly constrain the azeotropic behavior of the entrainer with the solutes.
An effective entrainer must enable separation by forming an appropriate azeotropic system, as discussed in \cite{Stichlmair1992}.
Enforcing a specific type of azeotropic behavior can be restrictive and difficult to formulate.
If a set of modelfluid features results in phase behavior that is incompatible with the separation task, the optimizer will be penalized by the violation of the product purity constraints.

\textit{(For brevity, the detailed bounds and specifications are omitted here but listed in \appendixref{app:optimizationdetails}, Step 2.)}

\begin{figure}[h]
    \centering
    \includegraphics[width=\textwidth]{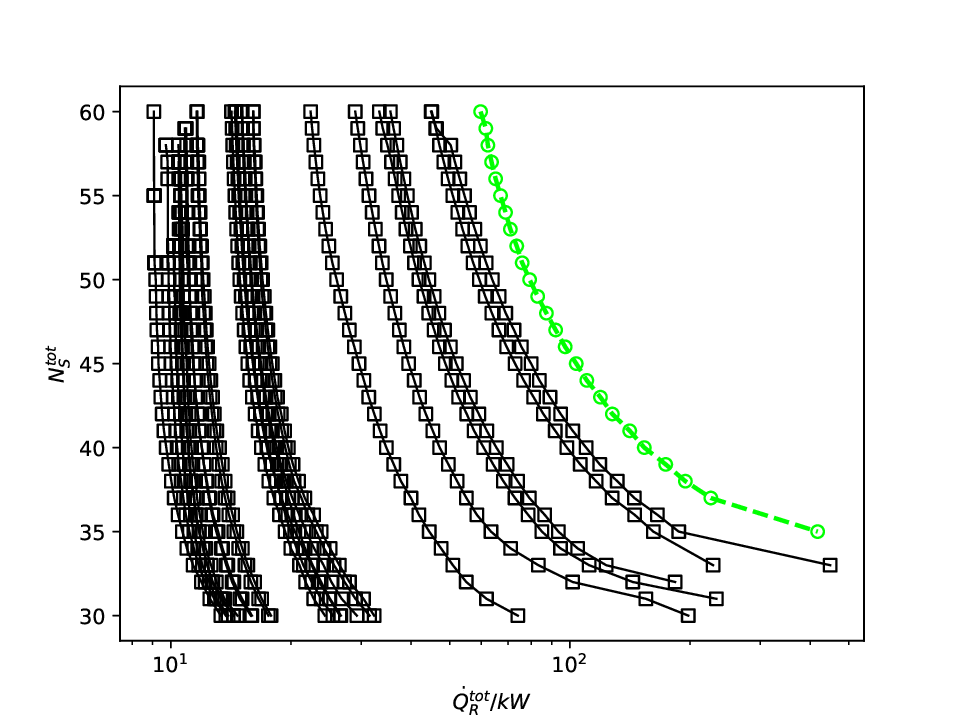}
    \caption{
        Comparison between the NQ curve obtained for the reference entrainer Benzene (lime-colored dash line with circle markers) and those obtained for the hypothetical optimal entrainers (black-colored solid lines with square markers) found in step 2.
        All NQ curves are computed by solving the optimization problem of step 1, but replacing the reference entrainer with the hypothetical entrainers found in step 2.
    }
    \label{fig:NQValidation:Hypo}
\end{figure}

As mentioned above, we use the Pareto-optimal solutions from Step 1 as starting points for this entrainer optimization.
For each point on the NQ curve, we fix the stages and run the optimization to find the corresponding hypothetical best entrainer.
The resulting hypothetically optimal entrainer components are compared to the reference entrainer in \figureref{fig:NQValidation:Hypo}.
The comparison is done by computing an NQ curve for each of the hypothetical entrainers found in step 2.
This means, we replace the reference entrainer in the optimization problem of step 1 by each of the hypothetical entrainers found in step 2, and solving it again, yielding a new NQ curve for each hypothetical entrainer.

\subsection{From Hypothetical Target to Real Candidate} \label{sec:entraineravailable}
The continuous optimization approach yields one or more hypothetically optimal entrainers, a set of ideal properties that may not correspond to any real molecule.
Compared to the alternative access to process fluid optimization, which is the direct optimization of the molecular structure, the continuous optimization approach comes with the additional challenge of mapping from the hypothetical to a real, available component.
In the field of process fluid optimization, to our best knowledge, this challenge was first addressed by \cite{Bardow2010}, who proposed a mapping strategy using candidate features and objective prediction.
Such mapping processes aim to identify the best-performing candidate from a given pool, which is not necessarily the one with properties closest to the hypothetical target, but rather the one predicted to yield the best process performance.
In this section we describe how the information on the hypothetically optimal entrainer candidate(s) can be used to approximate the efficiency of an arbitrary entrainer candidate.

\paragraph{An ML-fueled candidate pool for entrainer selection}
In practice, there is another challenge in addition to that of identifying the best performing candidate: It is that of getting a sufficiently large candidate pool.
If the pool of candidates is small, the chances are small that there are many high-performing candidates in it.

For the example of entrainer selection, we either need to know the modelfluid features or thermodynamic model parameters for each of the candidates we want to consider.
Since the availability of pure component property (PCP) data usually is not a limiting factor, we use parameters for the Antoine vapor pressure equation available in the DIPPR 801 database \cite{Wilding1998}.
Also the availability of vaporization enthalpies at the saturated vapor temperature of the components candidates is not a limiting factor in this study.
However, the situation looks different for the activity coefficients at infinite dilution, for which data is scarce.
This is where we employ the machine learning prediction methods.
We use the Matrix-Completion-Method, as proposed in \cite{Jirasek2020} to predict the activity coefficients at infinite dilution for each binary mixture formed by all unique binary combinations of components in our database.
While in practice, one would only use such prediction methods if no existing parameters of established thermodynamic models (e.g. NRTL, SAFT, UNIFAC) are available for the present mixture or component, we use it for all entrainer candidates considered in this study for demonstration purposes.
This significantly increases the number of components that we can consider in the study on finding a suitable entrainer candidate for the separation of the Acetone + Benzene mixture.

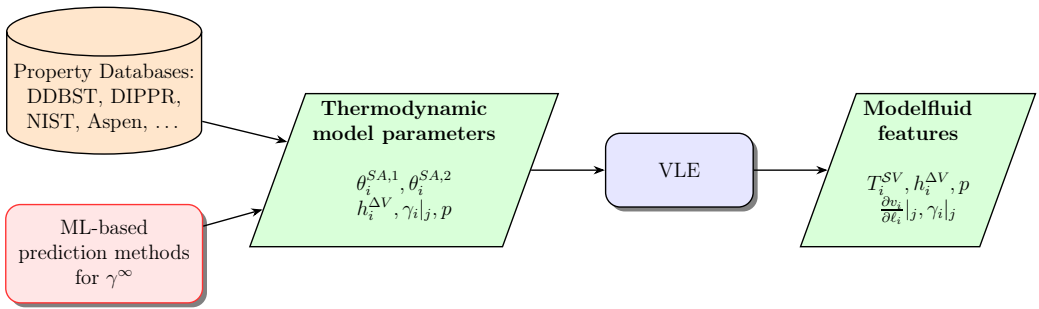
\begin{figure*}[t!]

\centering
\begin{adjustbox}{width=\textwidth, center}  
\begin{tikzpicture}[
    node distance=10mm and 15mm,
    process/.style={
        rectangle, 
        rounded corners=3mm, 
        draw, 
        thick, 
        fill=blue!10, 
        align=center,
        minimum height=15mm, 
        minimum width=30mm,
        drop shadow
    },
    data/.style={
        trapezium, 
        trapezium left angle=70, 
        trapezium right angle=110,
        draw, 
        thick, 
        fill=green!15, 
        align=center, 
        minimum height=10mm
    },
    database/.style={
        cylinder, 
        shape border rotate=90, 
        aspect=0.25, 
        draw, 
        thick, 
        fill=orange!20,
        align=center, 
        minimum height=15mm, 
        minimum width=25mm
    },
    digitaltwin/.style={
        rectangle, 
        rounded corners=3mm, 
        draw=red!80, 
        thick, 
        fill=red!10, 
        align=center,
        minimum height=20mm, 
        minimum width=40mm,
        drop shadow
    },
    arrow/.style={
            -Stealth,
            thick
        },
    label/.style={
        text=blue!80!black, 
        font=\small\bfseries
    },
    fitbox/.style={
        rectangle,
        rounded corners=5mm,
        draw=gray!50,
        thick,
        dashed
    }
]

\node (pcp_databases) [database] {Property Databases: \\ DDBST, DIPPR, \\ NIST, Aspen, \dots};
\node (ml_mc) [digitaltwin, below=of pcp_databases] {ML-based \\ prediction methods \\ for $\gamma^\infty$};
\node (parameters) [data, right=of ml_mc, yshift=17mm] {\textbf{Thermodynamic} \\ \textbf{model parameters} \\ \\ $\parameter_i^{\simplifiedantoine, 1}, \parameter_i^{\simplifiedantoine, 2}$ \\ $\vaporizationenthalpyof{i}, \activitycoefficientatinfinitedilutionin{i}{j}, \pressure$ \\};
\draw[arrow] (pcp_databases) -- (parameters);
\draw[arrow] (ml_mc) -- (parameters);
\node (VLE) [process, right=of parameters] {VLE};
\draw[arrow] (parameters) -- (VLE);
\node (features) [data, right=of VLE] {\textbf{Modelfluid} \\ \textbf{features} \\ \\ $\saturatedvaportemperatureof{i}, \vaporizationenthalpyof{i}, \pressure$ \\ $\derivativefeatureatinfdilution{i}{j}, \activitycoefficientatinfinitedilutionin{i}{j}$ \\};
\draw[arrow] (VLE) -- (features);

\end{tikzpicture}
\end{adjustbox}
\caption{
    Workflow of how the modelfluid features for each of the entrainer candidates is obtained using a PCP model parameter database and predictions of the activity coefficients at infinite dilution.
    The $\activitycoefficient_i\vert_j$ predictions are obtained using the Matrix Completion Method as proposed in \cite{Jirasek2020}.
}
    \label{fig:entrainercandidategeneration}
\end{figure*}

In \figureref{fig:entrainercandidategeneration}, we visualize the workflow of how to obtain the modelfluid features for each entrainer candidate in our pool.
First, the thermodynamic model parameters for each candidate are obtained by combining the PCP model parameters from the DIPPR database with the activity coefficients at infinite dilution, predicted by the ML methods.
Then, the modelfluid features are obtained from a VLE computed using these thermodynamic model parameters.

\paragraph{Candidate Filtering}
As we now have a large ML-fueled candidate pool, it is efficient to pre-screen the database of potential solvents and filter candidates not suitable for the separation task.
This filtering step follows established thermodynamic heuristics for entrainer selection, as outlined by works like \cite{Brignole1986} and \cite{Stichlmair1992, Duessel1995}.
For the separation of a maximum-boiling azeotrope, we consider a component a suitable entrainer if it meets one of the following criteria for homogeneous systems:
\begin{enumerate}
    \item \textbf{High-boiling entrainer:} The entrainer is the highest-boiling component in the ternary mixture.
    \item \textbf{Medium-boiling entrainer:} The entrainer's boiling point is between those of the two original components, and it forms a new maximum-boiling azeotrope with the highest-boiling original component.
    \item \textbf{Low-boiling entrainer:} The entrainer is the lowest-boiling component and forms new maximum-boiling azeotropes with both original components, at least one of which boils at a higher temperature than the original azeotrope.
\end{enumerate}
The above criteria are used in the subsequent paragraph when validating the mapping strategy-based rating of entrainer candidates.

\paragraph{Mapping Strategy via Objective Function Prediction}
To identify the most promising real candidates, we adopt a mapping strategy based on a local approximation of the objective function around the optimal hypothetical point.
This concept follows the work of \cite{Stavrou2023}, who use second-order Taylor expansions for this purpose.
In this work, we take advantage of the direct relationship between the modelfluid features and the distillation process, which allows us to predict the performance of real candidates based on their modelfluid features. For our case study, this means predicting the total heat duty based on the entrainer's features $\totalreboilerduty\inb{\feature_{\entrainer}} \approx \hat{\heatduty}^{\reboiler, \total}\inb{\feature_{\entrainer}, \feature_{\entrainer, \process}^{\best}}$:
\begin{equation} \label{eq:opexprediction}
    \hat{\heatduty}^{\reboiler, \total}\inb{\feature_{\entrainer}, \feature_{\entrainer, \process}^{\best}} = \totalreboilerduty\inb{\feature_{\entrainer, \process}^{\best}} + \nabla_{\feature_{\entrainer}} \reboilerduty\vert_{\feature_{\entrainer, \process}^{\best}} \cdot \inb{\feature_{\entrainer} - \feature_{\entrainer}^{\best}} .
\end{equation}
The gradient $\nabla_{\feature_{\entrainer}} \reboilerduty$ is readily available from the solution of the hypothetical entrainer optimization problem.
Since, at the location of the hypothetical optimal entrainer $\feature_{\entrainer, \process}^{\best}$, the gradient of the Lagrangian of the optimization problem \eqref{eq:entraineroptim:objectivefunction} is zero due to active constraints, the gradient of the objective function $\nabla_{\feature_{\entrainer}{\best}} \reboilerduty$ (without the constraints and Lagrangian multipliers) is non-zero.

Another difference of the mapping employed here to the work of Stavrou et al. \cite{Stavrou2023}, who use second-order Taylor expansions of both, the objective function and the constraints, is that we only consider a Taylor expansion of the objective function in this work.
While that may lead to small constraint violations when taking a step in the direction of the gradient, we argue that this is acceptable in our case, considering the uncertainties already present in the process.

As discussed in step 2 of \sectionref{sec:entrainerhypothetical}, we obtain a set of 23 individual optimal entrainers.
Since each of them is obtained as the solution to an individual optimization problem (different column sizes and feed positions for each of them), we refrain from selecting a single best hypothetical entrainer from this set.
Instead, we consider \eqref{eq:opexprediction} for each hypothetical optimal entrainer $\feature_{\entrainer,\process}^{\best}$ identified in step 2 of \sectionref{sec:entrainerhypothetical} and use its mean $\bar{\heatduty}^{\reboiler, \total}\inb{\feature_{\entrainer}}$
\begin{equation} \label{eq:meanopexprediction}
    \bar{\heatduty}^{\reboiler, \total}\inb{\feature_{\entrainer}} = \frac{1}{n} \sum_{i=1}^{n} \hat{\heatduty}^{\reboiler, \total} \inb{\feature_{\entrainer}, \feature_{\entrainer, \process}^{\best, i}} ,
\end{equation}
where $n=23$ is the number of hypothetical entrainers and $\feature_{\entrainer, \process}^{\best, i}$ is the $i$-th hypothetical optimal entrainer found in step 2 of \sectionref{sec:entrainerhypothetical} for some number of total stages in the flowsheet.

The workflow of using the set of hypothetical optimal entrainers to predict the efficiency of a solvent candidate in the entrainer distillation process is visualized in \figureref{fig:entrainermapping}.
\begin{figure*}[t!]

\centering
\begin{adjustbox}{width=\textwidth, center}  
\begin{tikzpicture}[
    node distance=10mm and 15mm,
    process/.style={
        rectangle, 
        rounded corners=3mm, 
        draw, 
        thick, 
        fill=blue!10, 
        align=center,
        minimum height=15mm, 
        minimum width=30mm,
        drop shadow
    },
    data/.style={
        trapezium, 
        trapezium left angle=70, 
        trapezium right angle=110,
        draw, 
        thick, 
        fill=green!15, 
        align=center, 
        minimum height=10mm
    },
    database/.style={
        cylinder, 
        shape border rotate=90, 
        aspect=0.25, 
        draw, 
        thick, 
        fill=orange!20,
        align=center, 
        minimum height=15mm, 
        minimum width=25mm
    },
    digitaltwin/.style={
        rectangle, 
        rounded corners=3mm, 
        draw=red!80, 
        thick, 
        fill=red!10, 
        align=center,
        minimum height=20mm, 
        minimum width=40mm,
        drop shadow
    },
    arrow/.style={
            -Stealth,
            thick
        },
    label/.style={
        text=blue!80!black, 
        font=\small\bfseries
    },
    fitbox/.style={
        rectangle,
        rounded corners=5mm,
        draw=gray!50,
        thick,
        dashed
    }
]

\node (pcp_databases) [database] {Property Databases: \\ DDBST, DIPPR, \\ NIST, Aspen, \dots};
\node (ml_mc) [digitaltwin, below=of pcp_databases] {ML-based \\ prediction methods \\ for $\gamma^\infty$};
\node (phase1_box) [fitbox, fit=(pcp_databases) (ml_mc), inner sep=8mm] {};
\node[label, above=3mm of phase1_box.north] {Entrainer candidates};

\node (hypofluid) [data, right=of phase1_box, yshift=18mm] {Multiple hypothetical \\ optimal entrainers for \\ different column sizes};
\node (opex_prediction) [digitaltwin, below=of hypofluid] {Compute 1\textsuperscript{st} order \\ Taylor series \\ approximation to \\ objective function \\ at optimal solution};
\draw[arrow] (hypofluid) -- (opex_prediction);
\node (hypo_pred_box) [fitbox, fit=(hypofluid) (opex_prediction), inner sep=8mm] {};
\node[label, above=1mm of hypo_pred_box.north] {Process Fluid Optimization};

\draw[arrow] (phase1_box) -- (hypo_pred_box);

\node (candidate) [data, right=of hypo_pred_box] {$\bar{\heatduty}^{\reboiler, \total}\inb{\feature_{\entrainer}}$ \\ \eqref{eq:meanopexprediction}};
\node (foreach_box) [fitbox, fit=(candidate), inner sep=8mm] {};
\node[label, above=1mm of foreach_box.north] {For each candidate};

\draw[arrow] (hypo_pred_box) -- (foreach_box);

\node (sorting_rule) [process, right=of foreach_box, yshift=-16mm] {Sort candidates \\ using their mean \\ estimated objective};
\node (sorted_order) [database, above=of sorting_rule] {Sorted \\ entrainer \\ candidates};
\draw[arrow] (sorting_rule) -- (sorted_order);
\node (sorting_box) [fitbox, fit=(sorting_rule) (sorted_order), inner sep=8mm] {};
\node[label, above=1mm of sorting_box.north] {Mean sorting};

\draw[arrow] (foreach_box) -- (sorting_box);

\end{tikzpicture}
\end{adjustbox}
    \label{fig:entrainermapping}
\end{figure*}

The usage of a local approximation on the efficiency of entrainer candidates in combination with the ML-fueled candidate pool enables the consideration of large candidate databases.
Given a modelfluid feature vector, the efficiency of an arbitrary entrainer component in the role as entrainer for the present separation task can be approximated using \eqref{eq:opexprediction}, without the necessity of any additional simulation or optimization.
The availability of the modelfluid feature vectors for large candidate pools is provided by the usage of the activity coefficient prediction methods.

\paragraph{Ranking and Validation}
Using \eqref{eq:opexprediction}, the candidates in the entrainer pool can be ranked according to their predicted performance.
Instead of selecting a single "best" candidate, we generate a ranked list from best to worst, a practice also advocated by \cite{Raman1998}.
This provides a more comprehensive overview and is more robust to potential inaccuracies in property predictions.
To validate the results presented in the following, a rigorous optimization was performed for each candidate, i.e. the NQ curve was computed for each candidate using the process described in the above in step 1.
Due to the high computational cost involved in the generation of rigorous NQ curves for validation, the results presented in the following feature only $39$ entrainer candidates.
This is not a limitation of the proposed modelfluid-based approach to process fluid optimization but a technical limitation on the validation side.

The ranking is generated by applying \eqref{eq:meanopexprediction} to each candidate in the solvent pool, which yields an efficiency prediction for each candidate at a specific operating point.
Considering \eqref{eq:meanopexprediction} for each $\feature_{\entrainer, \process}^{\best}$ for the different numbers of stages in the flowsheet effectively yields a predicted NQ curve for each candidate in the entrainer pool.
For the ranking, we consider one candidate "better" than another if the majority of points on its NQ curve are below the other candidate's curve, indicating lower energy consumption over the considered numbers of stages.

To validate the accuracy of this ranking, we performed a rigorous but computationally intensive benchmark: a full NQ curve optimization was computed for every single candidate in the pool.
The significant computational cost of this enumeration—several hundred CPU hours—underscores the practical necessity of our workflow (\figureref{fig:entrainermapping}) for efficient decision-making.
%

\begin{figure}[h]
    \centering
    \includegraphics[width=0.7\textwidth]{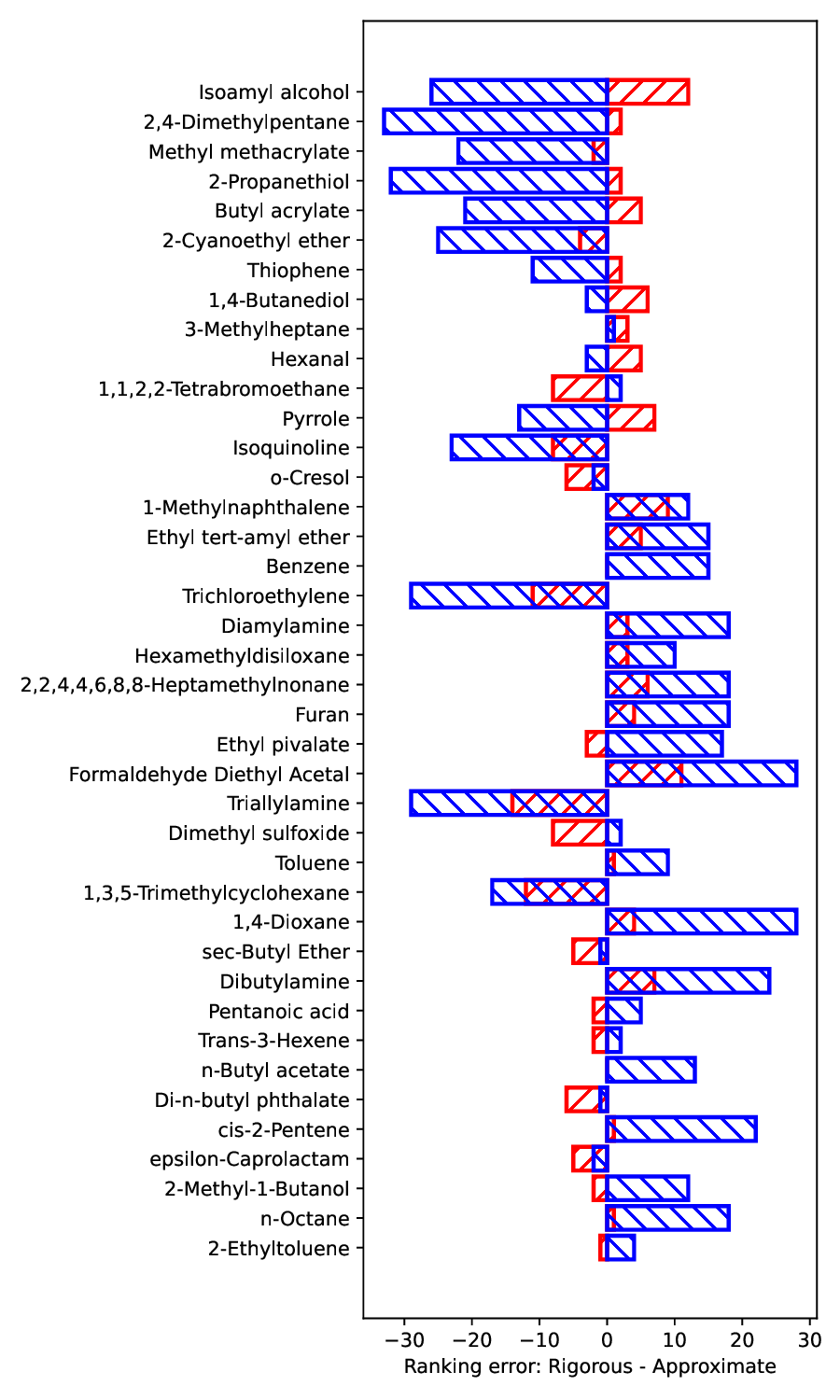}
    \caption{
        Quantitative sorting error.
        The horizontal axis shows the difference between the true rank and the predicted rank.
        Red bars ("//") show the error when ranking from the hypothetical optimum; blue bars ("\\") show the error when ranking from the reference entrainer.
        Red bars ("//") show the error when ranking from the hypothetical optimum; blue bars ("\textbackslash\textbackslash") show the error when ranking from the reference entrainer.
        The smaller red bars confirm the superior accuracy of our method.
    }
    \label{fig:SortingError}
\end{figure}

The results of this validation are presented in \figureref{fig:SortingError}, which enables for a quantitative assessment of the results.
The sorting error in \figureref{fig:SortingError} is defined as the difference between a candidate's predicted rank and its true rank determined from the validation (computation of the NQ curve for each candidate).
The entrainer candidates are sorted from best (bottom) to worst (top) along the vertical axis of \figureref{fig:SortingError}.
From the red bars in \figureref{fig:SortingError}, it can be seen that the sorting error is particularly small in the vicinity of the best candidates (bottom) and near the reference entrainer (Benzene).
We enrich that comparison by a second ranking: Instead of using the hypothetical optimum entrainer- and process features in \eqref{eq:meanopexprediction}, we also compute the ranking using the reference entrainer's fluid and process features in \eqref{eq:opexprediction}.
The chart critically compares the ranking generated from our hypothetical optimum (red bars) against a ranking generated from the reference entrainer, Benzene (blue bars).
The significantly smaller error for the red bars demonstrates that performing the entrainer optimization to find the hypothetical target is a crucial step for accurate candidate ranking.
In particular, the most promising candidates are positioned with high accuracy, proving the value of our approach.

\FloatBarrier

Finally, a detailed analysis of the modeling and simulation errors inherent in our approach is provided in \appendixref{sec:modelingerror} and \appendixref{sec:simulationerror}.
The analysis shows that these errors are generally small and do not significantly impact the validity of the sorting order presented here.
Therefore, while it is possible to incorporate error analysis into subsequent studies, it is not essential for the conclusions drawn in this section.

\section{Conclusion and Outlook} \label{sec:conclusion}
This work introduces a novel modelfluid representation tailored for distillation-based processes and demonstrates its utility for integrated process and entrainer design.
We developed a set of physically-meaningful features derived directly from vapor liquid equilibrium (VLE) data.
These features, which have an explainable effect on distillation performance, can be mapped to the parameters of simple thermodynamic models, enabling simple integration in existing process simulation tools.
We applied this framework to the entrainer selection problem for a complex distillation flowsheet, performing a continuous optimization of the entrainer's modelfluid features to identify hypothetically optimal candidates.

The approach successfully identified the most promising candidates from an entrainer pool.
Using a simple first-order Taylor approximation around the hypothetical optimum, we generated a ranked list of real candidates.
This ranking achieved a strong correlation with the true performance order, which was established through a computationally exhaustive validation involving rigorous NQ optimizations for every candidate.
Crucially, we demonstrated that finding the hypothetical optimum first is essential for this accuracy; a ranking generated from the reference entrainer's location yielded significantly poorer results, proving that proximity to the true optimum is vital for a reliable local approximation in such a complex flowsheet.
The success of the simple linear approximation further validates the design of our modelfluid, whose features have a more direct impact on process performance than more abstract parameters.

The broader impact of this work lies in its ability to bridge missing data in process design studies.
We show that our framework can be powerfully combined with modern ML-based property prediction methods, such as the Matrix Completion technique for activity coefficients at infinite dilution \cite{Jirasek2020}.
This synergy enables the consideration of vast candidate pools where experimental data is sparse, tackling a major bottleneck in solvent selection.
In doing so, our methodology provides a high-level framework for optimal experimental design.
While traditional model-based optimal experimental design \cite{Bubel2024} is highly data-efficient, it remains infeasible for screening the high number of potential mixtures in entrainer selection.
Our method offers a strategic pre-selection, guiding limited experimental resources toward the most promising candidates first.

Finally, we envision that custom, physically-meaningful modelfluid representations like the one presented here can serve as a superior basis for the training of explicit surrogate models.
By learning process behavior in this more intuitive feature space, surrogate models could achieve the generalization and robustness needed to accelerate their transition from academic prototypes to powerful tools for industrial process design and optimization.

\appendix
\counterwithin{equation}{section}

\section{Feature reduction for multi-component systems} \label{sec:modelfluidparameters:reduction}

This section provides detailed derivation of the feature reduction that occurs when moving from a binary to a multi-component system, where the number of independent modelfluid features does not simply scale with the number of new component pairs.
Instead, a feature reduction occurs due to the intrinsic physical properties of the pure components.
This section details the mathematical origin of this reduction.

The core principle is that the properties of a pure component, such as its vapor pressure as a function of temperature, are inherent to that component.
These properties remain consistent regardless of the other components with which it is mixed.
In our framework, this means the simplified Antoine parameters (\(\parameter_i^{\simplifiedantoine, 1}\), \(\parameter_i^{\simplifiedantoine, 2}\)) for component \(i\) are constant, whether it is part of a binary system with component \(j\) or a different binary system with component \(k\).

We can formalize this by considering the vapor liquid equilibrium, governed by the extended Raoult's Law:
\begin{equation} \label{eq:extendedraoultslaw}
    \vapormolarfractionof{i} \cdot \pressure = \liquidmolarfractionof{i} \cdot \activitycoefficient_i\inb{\liquidmolarfractions, \temperature} \cdot \pressure_{i}^{\saturatedvapor}\inb{\temperature}
\end{equation}
A key feature in our model is the slope of the VLE curve at infinite dilution.
Its definition follows from the derivative of \eqref{eq:extendedraoultslaw} with respect to the liquid molar fraction of component \(i\)
\begin{align} \label{eq:extendedraoultslawderivative}
    &\frac{\partial \vapormolarfractionof{i}}{\partial \liquidmolarfractionof{i}} \cdot \pressure = \pressure_{i}^{\saturatedvapor}\inb{\temperature} \cdot \inb{\activitycoefficient_i\inb{\liquidmolarfractions, \temperature} + \liquidmolarfractionof{i} \cdot \frac{\partial \activitycoefficient_i\inb{\liquidmolarfractions, \temperature}}{\partial \liquidmolarfractionof{i}}} \\  \nonumber
    &\text{at infinite dilution in component} j \text{:} \\ \nonumber
    &\frac{\partial \vapormolarfractionof{i}}{\partial \liquidmolarfractionof{i}}\vert_j \cdot \pressure = \pressure_{i}^{\saturatedvapor}\inb{\temperature\vert_j} \cdot \activitycoefficient_i\inb{\liquidmolarfractions\vert_j, \temperature\vert_j} .
\end{align}

To demonstrate the consequence of the invariant pure component properties, we consider a ternary system with components 1, 2, and 3. We examine component 1 at infinite dilution in component 2, and separately, in component 3.
By taking the logarithm of the VLE relationship at these limits and incorporating the simplified Antoine equation for vapor pressure, we arrive at the following expressions:
\begin{align}
    \ln\inb{\frac{\partial \vapormolarfractionof{1}}{\partial \liquidmolarfractionof{1}}\vert_2} + \ln\inb{\pressure} &= \ln\inb{\activitycoefficient_1}\vert_2 + \parameter_1^{\simplifiedantoine, 1} + \frac{\parameter_1^{\simplifiedantoine, 2}}{\saturatedvaportemperatureof{2}} \label{eq:log_relation_1_in_2} \\
    \ln\inb{\frac{\partial \vapormolarfractionof{1}}{\partial \liquidmolarfractionof{1}}\vert_3} + \ln\inb{\pressure} &= \ln\inb{\activitycoefficient_1}\vert_3 + \parameter_1^{\simplifiedantoine, 1} + \frac{\parameter_1^{\simplifiedantoine, 2}}{\saturatedvaportemperatureof{3}} \label{eq:log_relation_1_in_3}
\end{align}
Here, \(\saturatedvaportemperatureof{j}\) is the boiling temperature of pure component \(j\) at the system pressure \(\pressure\), and the terms on the left are features of the binary subsystems (1-2 and 1-3).
The parameters \(\parameter_1^{\simplifiedantoine, 1}\) and \(\parameter_1^{\simplifiedantoine, 2}\) are constant as they describe pure component 1.

We use the definition of the first Antoine parameter for a pure component,
\begin{equation}
    \parameter_i^{\simplifiedantoine, 1} = \ln\inb{\pressure} - \frac{\parameter_i^{\simplifiedantoine, 2}}{\saturatedvaportemperatureof{i}}
\end{equation}
to substitute for \(\parameter_1^{\simplifiedantoine, 1}\) in \eqref{eq:log_relation_1_in_2} and \eqref{eq:log_relation_1_in_3}.
After rearranging both equations to solve for the second Antoine parameter, \(\parameter_1^{\simplifiedantoine, 2}\), we obtain:
\begin{align}
    \parameter_1^{\simplifiedantoine, 2} \cdot \inb{\frac{1}{\saturatedvaportemperatureof{2}} - \frac{1}{\saturatedvaportemperatureof{1}}} &= \ln\inb{\frac{\partial \vapormolarfractionof{1}}{\partial \liquidmolarfractionof{1}}\vert_2} - \ln\inb{\activitycoefficient_1}\vert_2 \\
    \parameter_1^{\simplifiedantoine, 2} \cdot \inb{\frac{1}{\saturatedvaportemperatureof{3}} - \frac{1}{\saturatedvaportemperatureof{1}}} &= \ln\inb{\frac{\partial \vapormolarfractionof{1}}{\partial \liquidmolarfractionof{1}}\vert_3} - \ln\inb{\activitycoefficient_1}\vert_3
\end{align}
These two equations provide independent routes to calculate the same intrinsic property, \(\parameter_1^{\simplifiedantoine, 2}\).
Isolating this parameter reveals that the terms describing the system behavior must be equal:
\begin{align} \label{eq:modelfluidmappingdependency}
    \parameter_1^{\simplifiedantoine, 2} &= \left[\ln\inb{\frac{\partial \vapormolarfractionof{1}}{\partial \liquidmolarfractionof{1}}\vert_2} - \ln\inb{\activitycoefficient_1}\vert_2\right] \cdot \left[\frac{1}{\saturatedvaportemperatureof{2}} - \frac{1}{\saturatedvaportemperatureof{1}}\right]^{-1} \\
    \nonumber
    &= \left[\ln\inb{\frac{\partial \vapormolarfractionof{1}}{\partial \liquidmolarfractionof{1}}\vert_3} - \ln\inb{\activitycoefficient_1}\vert_3\right] \cdot \left[\frac{1}{\saturatedvaportemperatureof{3}} - \frac{1}{\saturatedvaportemperatureof{1}}\right]^{-1}
\end{align}
\eqref{eq:modelfluidmappingdependency} establishes a mathematical constraint linking the features of the 1-2 subsystem to the features of the 1-3 subsystem.

This same logic applies to component 2 (in solvents 1 and 3) and component 3 (in solvents 1 and 2), generating a total of three such constraints for a ternary system.
Each constraint removes one degree of freedom from the model.
This effect does not occur in a binary system, where no such shared-component dependency exists. For the ternary case, these three constraints reduce the number of independent features, yielding the 16 modelfluid features.

\section{Thermodynamic Feasibility Constraints} \label{sec:thermodynamicfeasibility}
While our modelfluid representation allows for a broad exploration of the chemical design space, not all combinations of feature values correspond to physically realistic fluid behavior.
To guide the optimization towards meaningful and achievable solutions, we introduce a set of thermodynamic feasibility constraints.
These constraints act as guardrails, preventing the optimizer from exploring highly artificial regions of the feature space that cannot be mapped back to real candidate molecules.
We formulate three key checks based on established thermodynamic principles.

First, we enforce liquid phase stability. A mixture must remain a single liquid phase to be thermodynamically stable.
This is ensured if the Gibbs excess energy function is convex, a condition we can evaluate using our activity coefficient model. For a binary mixture, this simplifies to the following condition on one of the components \cite[p. 505]{Smith1996}:
\begin{equation} \label{eq:thermoconstraint:phasestabilitycondition}
\left[\frac{\partial \ln\inb{\activitycoefficient_i\inb{\liquidmolarfractionof{i}, \temperature}}}{\partial \liquidmolarfractionof{i}}\right]_{\temperature, \pressure} + \frac{1}{\liquidmolarfractionof{i}} > 0 \quad \forall \liquidmolarfractionof{i} \in \left(0, 1\right) .
\end{equation}
Due to the Gibbs-Duhem relation, this single check is sufficient for a binary system. This stability test is a common feature in PFO literature, with similar conditions used in \cite{Buxton1999} and \cite{Hugo2004}.
Our full derivation for multi-component systems is provided in \appendixref{sec:phasestabilityderivation}.
This derivation, which aligns with \cite[p. 503f]{Smith1996}, also corrects a sign error present in the final stability equation reported by \cite{Hugo2004}.

Second, we consider the condition for azeotrope formation, a critical phenomenon in distillation.
For a binary system at constant pressure, a homogeneous azeotrope exists if the boiling temperature curve exhibits an extremum.
This is true if the slope of the curve at the pure component limits have opposite signs \cite{Buxton1999, Hugo2004}:
\begin{equation}
\frac{\partial \temperature}{\partial \liquidmolarfractionof{i}}\vert_{x_i=1} \cdot \frac{\partial \temperature}{\partial \liquidmolarfractionof{i}}\vert_{x_i=0} < 0 .
\end{equation}
We derive this condition from Raoult's Law in \appendixref{sec:isobaricazeotropyderivation}, where we also discuss its extension to differentiate between maximum- and minimum-boiling azeotropes.
We acknowledge that this condition is limited to detecting azeotropes in binary pairs within a multi-component system.

Third, we constrain the system to remain below its critical point.
Since the modelfluid is conceptualized from vapor liquid equilibrium data, its predictions become invalid in the supercritical region.
To enforce this without adding the complexity of a full equation of state to our modelfluid, we adopt a well-known heuristic from \cite[p. 16]{Poling2001} to approximate the critical temperature:
\begin{equation} \label{eq:criticaltemperatureapproximation}
\temperature_i^{\critical} \approx 1.5 \cdot \temperature_i^{\saturatedvapor} ,
\end{equation}
where $\temperature_i^{\critical}$ is the critical temperature and $\temperature_i^{\saturatedvapor}$ is the saturated vapor temperature of component $i$.
This simple rule ensures that the process operates in the subcritical domain, keeping the optimized fluid properties within a plausible physical range.
These three constraints are essential for the entrainer selection study in this work.
By embedding them within the optimization problem, we ensure that the search for the optimal entrainer is confined to a space of thermodynamically feasible candidates.
This is critical for producing an optimal theoretical result that is not unnecessarily far from the properties of available, real-world molecules, thereby making the subsequent candidate screening more effective.

\subsection{Phase stability condition} \label{sec:phasestabilityderivation}

The thermodynamic criterion for a liquid mixture to be stable as a single phase, at constant temperature and pressure, is that its Gibbs energy of mixing, \(\Delta G\), must be at a global minimum.
This implies that the function \(\Delta G(\liquidmolarfractionof{1}, \dots, \liquidmolarfractionof{N-1})\) must be convex.
For a binary system, this simplifies to the well-known condition that the second derivative of \(\Delta G\) with respect to composition must be positive across the entire composition range \cite[p. 504f]{Smith1996}:
\begin{equation} \label{eq:phasestabilitycondition}
    \left[\frac{\partial^2 \Delta G}{\partial \liquidmolarfractionof{i}^2}\right]_{\pressure, \temperature} > 0 \quad \forall \liquidmolarfractionof{i} \in (0, 1) .
\end{equation}
If this condition is not met, the system can lower its total Gibbs energy by splitting into two distinct liquid phases.

\subsubsection{Derivation for Binary Systems}
The Gibbs energy of mixing, \(\Delta G\), is defined as the difference between the Gibbs energy of the mixture, \(G\), and the mole-fraction-weighted sum of the pure component Gibbs energies, \(G_i\):
\begin{equation} \label{eq:mixinggibbsenergy}
    \Delta G = G - \sum_{i=1}^{N} \liquidmolarfractionof{i} \cdot G_i .
\end{equation}
The Gibbs energy of a non-ideal mixture is given by \cite[p. 390]{Smith1996}:
\begin{equation} \label{eq:mixturegibbsenergy}
    G = \sum_{i=1}^{N} \liquidmolarfractionof{i} G_i + R \temperature \sum_{i=1}^{N} \liquidmolarfractionof{i} \ln\inb{\liquidmolarfractionof{i}} + G^E ,
\end{equation}
where \(G^E\) is the excess Gibbs energy, which accounts for non-ideal interactions. Substituting \eqref{eq:mixturegibbsenergy} into \eqref{eq:mixinggibbsenergy} yields the common expression for the Gibbs energy of mixing:
\begin{equation}
    \Delta G = R \temperature \sum_{i=1}^{N} \liquidmolarfractionof{i} \ln\inb{\liquidmolarfractionof{i}} + G^E.
\end{equation}
The excess Gibbs energy is directly related to the activity coefficients, \(\activitycoefficient_i\), by \cite[p. 351]{Smith1996}:
\begin{equation} \label{eq:excessgibbsenergy}
    G^E = R \temperature \sum_{i=1}^{N} \liquidmolarfractionof{i} \ln\inb{\activitycoefficient_i}.
\end{equation}
By applying the stability criterion from \eqref{eq:phasestabilitycondition} and using the relationship between \(G^E\) and \(\activitycoefficient_i\), one can derive the stability condition in terms of the activity coefficient.
As shown in detail by Smith et al. \cite[p. 505]{Smith1996}, this results in:
\begin{equation} \label{eq:phasestabilityconditionmixingactivitycoefficient}
    \left[\frac{\partial \ln\inb{\activitycoefficient_i}}{\partial \liquidmolarfractionof{i}} + \frac{1}{\liquidmolarfractionof{i}}\right]_{\temperature, \pressure} > 0 \quad \forall \liquidmolarfractionof{i} \in (0, 1) .
\end{equation}
For a binary system, this condition only needs to be checked for one component, as the Gibbs-Duhem equation ensures it will also hold for the other.

\subsubsection{Generalization to Multicomponent Systems}
For a multicomponent mixture with \(N-1\) independent mole fractions, the stability criterion in \eqref{eq:phasestabilitycondition} generalizes to the requirement that the Hessian matrix of \(\Delta G\) must be positive definite.
This is equivalent to the condition that the determinant of the Hessian matrix must be positive \cite{Wisniak1983}:
\begin{equation}
    \det \left|
    \begin{bmatrix}
        \frac{\partial^2 \Delta G}{\partial \liquidmolarfractionof{1}^2} & \frac{\partial^2 \Delta G}{\partial \liquidmolarfractionof{1} \partial \liquidmolarfractionof{2}} & \dots & \frac{\partial^2 \Delta G}{\partial \liquidmolarfractionof{1} \partial \liquidmolarfractionof{N-1}} \\
        \vdots & \ddots &  & \vdots \\
        \frac{\partial^2 \Delta G}{\partial \liquidmolarfractionof{N-1} \partial \liquidmolarfractionof{1}} & \dots & \dots & \frac{\partial^2 \Delta G}{\partial \liquidmolarfractionof{N-1}^2}
    \end{bmatrix}_{\pressure, \temperature}
    \right| > 0.
\end{equation}
Since this work deals with ternary mixtures at most (\(N=3\)), we have \(N-1=2\) independent mole fractions.
The stability criterion simplifies to the determinant of a 2x2 matrix:
\begin{equation} \label{eq:ternaryphasestabilitycondition}
    \left[\frac{\partial^2 \Delta G}{\partial \liquidmolarfractionof{1}^2} \cdot \frac{\partial^2 \Delta G}{\partial \liquidmolarfractionof{2}^2} - \left(\frac{\partial^2 \Delta G}{\partial \liquidmolarfractionof{1} \partial \liquidmolarfractionof{2}}\right)^2 \right]_{\pressure, \temperature} > 0.
\end{equation}

\subsubsection{Implementation in Numerical Optimization}
For use in constrained numerical optimization as described in \sectionref{sec:entrainersearch}, the strict inequality `$>$` is often relaxed.
The boundary case where the second derivative is zero corresponds to "incipient instability" \cite[p. 273]{Prausnitz1998}, where the system is on the verge of phase splitting.
For practical purposes, we consider this state stable and modify the condition to be non-negative.

For the binary case, the implemented constraint is:
\begin{equation} \label{eq:optim:phasestabilityconditionmixingactivitycoefficient}
    \left[\frac{\partial \ln\inb{\activitycoefficient_i}}{\partial \liquidmolarfractionof{i}} + \frac{1}{\liquidmolarfractionof{i}}\right]_{\temperature, \pressure} \geq \varepsilon \quad \forall \liquidmolarfractionof{i} \in [0, 1] .
\end{equation}
A similar relaxation is applied to the ternary condition in \eqref{eq:ternaryphasestabilitycondition}.
We choose a tolerance of \(\varepsilon = 0\), acknowledging that the simplifications within our modeling approach and the inherent error of the modelfluid representation (\appendixref{sec:modelingerror}) make a strict distinction between stable and incipiently unstable states less critical.

\subsection{Isobaric azeotropy constraint} \label{sec:isobaricazeotropyderivation}

To detect the presence of binary azeotropes, we employ a powerful criterion based on the work of \cite{Buxton1999}.
The central idea is that for a binary mixture at constant pressure, a homogeneous azeotrope must exist if the slope of the bubble-point temperature with respect to composition, \(\partial \temperature / \partial \liquidmolarfractionof{i}\), has opposite signs at the two extremes of the composition range (i.e., at infinite dilution, \(\liquidmolarfractionof{i} \to 0\) and \(\liquidmolarfractionof{i} \to 1\)).
If the curve starts by increasing from one side and decreasing from the other, it must have a minimum or maximum point in between.

While this condition is specific to binary systems, it provides critical information for our work by allowing us to characterize the behavior of the binary subsystems (1-2, 1-3, and 2-3) within a ternary mixture.

\subsubsection{Mathematical Formulation}
To find the required derivative, \(\partial \temperature / \partial \liquidmolarfractionof{i}\), we start from the system of vapor liquid equilibrium equations defined by the extended Raoult's Law \eqref{eq:extendedraoultslaw}.
For a generic component \(i\), we can write this as an implicit function \(F_i\) that is zero at equilibrium:
\begin{equation}
    F_i\inb{\liquidmolarfractionof{i}, \pressure, \vapormolarfractionof{i}, \temperature} = \vapormolarfractionof{i} \pressure - \liquidmolarfractionof{i} \activitycoefficient_i\inb{\liquidmolarfractions, \temperature} \pressure_i^{\saturatedvapor}\inb{\temperature} = 0.
\end{equation}
For a binary system, this gives a system of two equations, \(F = [F_1, F_2]^T = 0\). In an isobaric system, the temperature \(\temperature\) and vapor mole fraction \(\vapormolarfractionof{i}\) are implicit functions of the liquid mole fraction \(\liquidmolarfractionof{i}\).
We can find the derivative of these dependent variables with respect to the independent ones using the implicit function theorem. Let \(\tilde{y} = [\temperature, \vapormolarfractionof{i}]^T\) be the vector of dependent variables and \(\tilde{x} = [\pressure, \liquidmolarfractionof{i}]^T\) be the vector of independent variables.
The Jacobian matrix of the dependent variables is then given by:
\begin{equation} \label{eq:partialtemperature}
    \frac{\partial \tilde{y}}{\partial \tilde{x}} = - \left[\frac{\partial F}{\partial \tilde{y}}\right]^{-1} \frac{\partial F}{\partial \tilde{x}}.
\end{equation}
The term \(\partial \temperature / \partial \liquidmolarfractionof{i}\) is the top-right element of the resulting \(2 \times 2\) matrix \(\partial \tilde{y} / \partial \tilde{x}\).
This procedure allows for its analytical or numerical calculation.

\subsubsection{Conditions for Azeotrope Existence and Type}
With the derivative \(\partial \temperature / \partial \liquidmolarfractionof{i}\) in hand, the condition for the existence of a binary azeotrope is that the derivatives at the composition limits have opposite signs:
\begin{equation} \label{eq:optim:azeotropecondition}
    \left(\frac{\partial \temperature}{\partial \liquidmolarfractionof{i}}\right)_{\liquidmolarfractionof{i} \to 0} \cdot \left(\frac{\partial \temperature}{\partial \liquidmolarfractionof{i}}\right)_{\liquidmolarfractionof{i} \to 1} < 0.
\end{equation}
The type of azeotrope can be further classified by inspecting the signs of the individual derivatives:
\begin{itemize}
    \item \textbf{Maximum-boiling azeotrope:} The temperature rises from both pure components towards the azeotrope.
    \begin{equation} \label{eq:optim:maximumazeotropetype}
        \left(\frac{\partial \temperature}{\partial \liquidmolarfractionof{i}}\right)_{\liquidmolarfractionof{i} \to 0} > 0 \quad \text{and} \quad \left(\frac{\partial \temperature}{\partial \liquidmolarfractionof{i}}\right)_{\liquidmolarfractionof{i} \to 1} < 0.
    \end{equation}
    \item \textbf{Minimum-boiling azeotrope:} The temperature drops from both pure components towards the azeotrope.
    \begin{equation} \label{eq:optim:minimumazeotropetype}
        \left(\frac{\partial \temperature}{\partial \liquidmolarfractionof{i}}\right)_{\liquidmolarfractionof{i} \to 0} < 0 \quad \text{and} \quad \left(\frac{\partial \temperature}{\partial \liquidmolarfractionof{i}}\right)_{\liquidmolarfractionof{i} \to 1} > 0.
    \end{equation}
\end{itemize}
This approach is particularly suitable for the optimization in \sectionref{sec:entrainersearch} because it directly tests for the existence of an azeotrope, which is our primary concern.
This contrasts with methods like those in \cite{Wisniak1983}, which are more focused on classifying the type of an already-known azeotrope.

\subsubsection{Limitations of the Conditions}
This method has two important limitations. First, it is designed to detect only binary, homogeneous azeotropes.
Second, its reliability depends on the physical realism of the underlying VLE model.
During the exploration of the modelfluid feature space (e.g., in \sectionref{sec:entrainersearch}), some feature combinations may produce "unphysical" VLE curves with non-monotonic behavior near the pure components that does not correspond to a true azeotrope.
In such cases, the algebraic conditions may yield a false positive. This can happen if, for example, the activity coefficient model is poorly parameterized, leading to incorrect behavior at infinite dilution.
Awareness of this limitation is crucial when interpreting the results of optimizations that use this constraint.

\section{Modelfluid-tailored distillation column simulation} \label{sec:fakeenthalpysimulation}

For the extensive optimization studies in this work, a computationally efficient yet reasonably accurate distillation column simulation is required. Standard rigorous methods, such as the one described by \cite{Hoffmann2017}, involve nested fixed-point iterations for temperature on each stage, making them too slow for our purposes. Therefore, we developed a modified stage-to-stage simulation method tailored to the assumptions of our modelfluid framework (\sectionref{sec:modelfluid}).

The key modification is the assumption that molar enthalpies are independent of temperature. This simplification eliminates the need for the inner temperature-solving loop, drastically increasing computational speed while retaining the core stage-by-stage mass balance logic. The results obtained with this approach maintain reasonable accuracy compared to rigorous methods, as will be analyzed in \appendixref{sec:simulationerror}.

\subsection{Governing Assumptions}
Our simulation model is built upon the following set of assumptions:
\begin{itemize}
    \item Total vaporization in the reboiler and total condensation in the condenser.
    \item Adiabatic column operation (no heat exchange with the surroundings).
    \item Isobaric column operation (negligible pressure drop across stages).
    \item All streams leaving a stage are in thermodynamic equilibrium (saturated liquid and vapor).
    \item Molar enthalpies (liquid, vapor, and vaporization) are independent of temperature but dependent on composition. We set the reference liquid molar enthalpy to zero, \(\molarenthalpy^{\liquid} = 0\), so the vapor molar enthalpy equals the enthalpy of vaporization, \(\molarenthalpy^{\vapor} = \molarenthalpy^{\vaporization}\).
\end{itemize}

\subsection{Comparison to Constant Molar Overflow (CMO)}
To clarify the position of our method, we contrast it with the well-known McCabe-Thiele or Constant Molar Overflow (CMO) assumptions. The CMO concept simplifies the problem such that VLE is the only physical property required, assuming that for every mole of vapor that condenses, one mole of liquid vaporizes \cite{Mairhofer2023}. This implies that components have identical latent heats of vaporization \cite{Mathew2023}.

Our method is significantly less restrictive than CMO:
\begin{itemize}
    \item \textbf{Enthalpies:} We account for composition-dependent vaporization enthalpies, allowing components to have different latent heats. This means molar flow rates of liquid and vapor (\(\liquidflow^{(i)}, \vaporflow^{(i)}\)) are not constant and vary from stage to stage.
    \item \textbf{VLE:} We do not assume constant relative volatility; VLE is calculated on each stage using the full thermodynamic model.
    \item \textbf{Scope:} Our method is numerical, handles multi-component systems, and operates with a finite number of stages, where feed conditions directly influence the separation. This contrasts with the graphical, binary-focused McCabe-Thiele method, which often analyzes limiting cases (e.g., infinite reflux or stages).
\end{itemize}

\subsection{Numerical Algorithm}
The simulation solves for the steady-state profile of a column with specified feed conditions, geometry, and operating parameters (\(\spl, \refluxratio\)). It is formulated as a root-finding problem on the bottom composition, \(\liquidmolarfractions^{(\bottom)}\). The objective is to find the \(\liquidmolarfractions^{(\bottom)}\) that ensures the liquid and vapor compositions, when calculated from opposite ends of the column, match at the feed stage.

The problem is to solve \(\constraint_{\column} = 0\), where \(\constraint_{\column}\) is the difference between the vapor composition at the feed stage calculated by the downward sweep and the equilibrium vapor composition corresponding to the liquid calculated by the upward sweep:
\begin{align} \label{eq:fakeenthalpysimulation:loss}
    \constraint_{\column}\inb{\liquidmolarfractions^{(\bottom)}, \dots} = \vapormolarfractions^{(\feed)}_{\text{down}} - \vapormolarfractions\inb{\liquidmolarfractions^{(\feed)}_{\text{up}}, \pressure, \temperature^{(\feed)}}
\end{align}
In this equation, \(\liquidmolarfractions^{(\feed)}_{\text{up}}\) is the liquid composition on the feed stage calculated from the bottom up, and \(\vapormolarfractions^{(\feed)}_{\text{down}}\) is the vapor composition on the feed stage calculated from the top down. The notation \(\liquidmolarfractions^{\feed}\) denotes the composition of the external feed stream, whereas \(\liquidmolarfractions^{(\feed)}\) refers to the composition of the mixed liquid on the feed stage itself.

The column stages are numbered as follows:
\begin{align}
    \text{Bottom stage (Reboiler)} &: \mynum_{\stages}^{(\bottom)} = 0 \\
    \text{Feed stage} &: \mynum_{\stages}^{(\feed)} = \mynum_{\stages}^{\belowfeed} + 1 \\
    \text{Total column stages} &: \mynum_{\stages}^{(\total)} = \mynum_{\stages}^{\abovefeed} + \mynum_{\stages}^{\belowfeed} + 1 \\
    \text{Distillate stage (Condenser)} &: \mynum_{\stages}^{(\distillate)} = \mynum_{\stages}^{(\total)} + 1
\end{align}

\subsection{Detailed Calculation Procedures}
The following sections detail the upward and downward calculation sweeps that constitute one evaluation of the function \(\constraint_{\column}\).

\paragraph{Upward Calculation (Bottom to Feed Stage)}
This sweep calculates the liquid and vapor profiles from the reboiler (\(i=0\)) up to the feed stage.
\begin{center} \label{fig:distillationcolumn:upwardscontrolvolume}
    \begin{tikzpicture}
        \bottomstage{0}{0}
        \normalstage{i-1}{1}
        \feedstage{i}{2}
    \end{tikzpicture}
\end{center}
The bottom product flow rate is set by the split fraction: \(\liquidflow^{(\bottom)} = \liquidflow^{\feed} \cdot \spl\). For any stage \(i+1\) in the stripping section, the overall mass balance is \(\liquidflow^{(i+1)} = \vaporflow^{(i)} + \liquidflow^{(\bottom)}\). Due to total vaporization in the reboiler, the vapor leaving it has the same composition as the liquid inside: \(\vapormolarfractions^{(0)} = \liquidmolarfractions^{(\bottom)}\).

The simplified enthalpy balance around the reboiler control volume yields the vapor molar flow rate from stage \(i\):
\begin{equation}
    \vaporflow^{(i)} = \frac{\reboilerduty}{\sum_{j=1}^{\mynum_{\component}} \vapormolarfractionof{j}^{(i)} \cdot \molarenthalpy_j^{\vaporization}}
\end{equation}
With the molar flow rates known, the component mass balance for stage \(i+1\) yields the liquid composition:
\begin{equation}
    \liquidmolarfractionof{j}^{(i+1)} = \frac{\liquidflow^{(\bottom)} \cdot \liquidmolarfractionof{j}^{(\bottom)} + \vaporflow^{(i)} \cdot \vapormolarfractionof{j}^{(i)}}{\liquidflow^{(i+1)}}
\end{equation}
The vapor composition \(\vapormolarfractions^{(i+1)}\) is then found by a bubble-point calculation for the liquid \(\liquidmolarfractions^{(i+1)}\). This process is repeated iteratively up to the feed stage.

\paragraph{Downward Calculation (Condenser to Feed Stage)}
This sweep calculates the profiles from the condenser down to the feed stage.
\begin{center}
    \begin{tikzpicture} \label{fig:distillationcolumn:downwardscontrolvolume}
        \distillatestage{\mynum_{\stages}^{\total}+1}{2}
        \normalstage{i}{1}
        \feedstage{i-1}{0}
    \end{tikzpicture}    
\end{center}
The distillate flow rate is \(\liquidflow^{(\distillate)} = \liquidflow^{\feed} \cdot (1 - \spl)\), and the reflux flow is \(\liquidflow^{({\mynum_{\stages}^{\total}+1})} = \refluxratio \cdot \liquidflow^{(\distillate)}\). Due to total condensation, the compositions of the distillate, reflux, and top vapor are equal: \(\liquidmolarfractions^{(\distillate)} = \liquidmolarfractions^{({\mynum_{\stages}^{\total}+1})} = \vapormolarfractions^{(\mynum_{\stages}^{\total})}\).

The enthalpy balance over a control volume enclosing the condenser and the top stages gives the vapor molar flow leaving stage \(i-1\):
\begin{equation}
    \vaporflow^{(i-1)} = \frac{-\condenserduty}{\sum_{j=1}^{\mynum_{\component}} \vapormolarfractionof{j}^{(i-1)} \cdot \molarenthalpy_j^{\vaporization}}
\end{equation}
The corresponding liquid molar flow from stage \(i\) is found via a mass balance on the same control volume: \(\liquidflow^{(i)} = \vaporflow^{(i-1)} + \liquidflow^{(\distillate)}\). Finally, the component mass balance yields the vapor composition we need:
\begin{equation}
    \vapormolarfractionof{j}^{(i-1)} = \frac{\liquidflow^{(i)} \cdot \liquidmolarfractionof{j}^{(i)} - \liquidflow^{(\distillate)} \cdot \liquidmolarfractionof{j}^{(\distillate)}}{\vaporflow^{(i-1)}}
\end{equation}
This set of equations is solved iteratively for each stage, starting from the condenser and moving down to the feed stage.

\section{Discussion of Modeling and Simulation Error}

This section discusses the two primary sources of error in our methodology: the approximation introduced by the \textbf{modelfluid representation} and the error from the \textbf{modelfluid-tailored distillation column simulation}.
For each, we first derive an approximate analytic expression for the error using first-order Taylor series expansions, as both the fluid modeling and column simulation are nonlinear.
We then provide a numerical consideration to quantify these errors in practice.

\subsection{Error of Fluid Modeling} \label{sec:modelingerror}

\paragraph{Analytic View}
The first source of error arises from approximating a complex, "rigorous" fluid model with our simplified "modelfluid" representation.
The rigorous model, denoted by superscript $\rigorous$, typically consists of a detailed vapor pressure equation and an activity coefficient model with parameters $\parameter^{\rigorous}$.

Our process begins by computing the vapor liquid equilibrium for a given system using this rigorous model.
From this VLE data, we extract a set of six characteristic features, $\modelfluidfeatures=\modelfluidfeatures\inb{\parameter^{\rigorous}}$, which define the modelfluid.
These features are the pure component saturation temperatures ($\saturatedvaportemperatureof{i}$), infinite dilution activity coefficients ($\activitycoefficientatinfinitedilutionin{i}{j}$), and a derivative-based feature at infinite dilution ($\derivativefeatureatinfdilution{i}{j}$):
\begin{gather}
    \modelfluidfeatures = \left[\saturatedvaportemperatureof{1}, \saturatedvaportemperatureof{2}, \activitycoefficientatinfinitedilutionin{1}{2}, \activitycoefficientatinfinitedilutionin{2}{1}, \derivativefeatureatinfdilution{1}{2}, \derivativefeatureatinfdilution{2}{1}\right] \\
    \saturatedvaportemperatureof{i}: \pressure = \saturatedvaporpressurevariantaofat{i}{\temperature} \quad \forall i \in \{1,2\} \label{eq:feature_Tsat} \\
    \activitycoefficientatinfinitedilutionin{i}{j} = \activitycoefficientvariantaofat{i} \quad \forall i \in \{1,2\} \label{eq:feature_gamma_inf} \\
    \derivativefeatureatinfdilution{i}{j} = \activitycoefficientvariantaofat{i} \cdot \frac{\saturatedvaporpressurevariantaofat{i}{\temperature\inb{\liquidmolarfractionof{i}=0}}}{\pressure} \label{eq:feature_derivative}
\end{gather}
\textit{Note that in the following, we omit some of the arguments of the functions for notational convenience.}
Using these features, we can parameterize simplified thermodynamic models for vapor pressure and activity coefficient, as described in \sectionref{sec:modelfluidparameters}.
The parameters to these thermodynamic models are obtained as an explicit, though nonlinear expression of the features: $\modelfluidparameters = \parameter\inb{\modelfluidfeatures}$.
Furthermore, the features $\modelfluidfeatures$ are obtained as a nonlinear function of the rigorous parameters $\parameter^{\rigorous}$ (VLE computations incl. Raoult's Law).
Consequently, the overall mapping from $\parameter^{\rigorous}$ to $\modelfluidparameters$ is nonlinear:
\begin{equation}
    \modelfluidparameters = \parameter\inb{\modelfluidfeatures\inb{\parameter^{\rigorous}}} .
\end{equation}

To quantify the resulting modeling error, we compare the VLE predictions from both models.
The governing equation is the extended Raoult's Law, which we write as an implicit function $F$ that must equal zero at equilibrium for a given liquid composition $\liquidmolarfractions$ and pressure $\pressure$:
\begin{equation} \label{eq:implicitraoultslaw}
    F\inb{\vapormolarfractions, \temperature} = \sum_{i=1}^{N_C} \left( \vapormolarfractionof{i} \pressure - \liquidmolarfractionof{i} \activitycoefficient_i \pressure_{i}^{\saturatedvapor} \right) = 0 .
\end{equation}
The specific forms for the rigorous and modelfluid models are:
\begin{align}
    \label{eq:implicitraoultslaw:rigorous}
    F^{\rigorous} &= \sum_{i=1}^{N_C} \left( \vapormolarfractionof{i} \pressure - \liquidmolarfractionof{i} \activitycoefficient_i^{\rigorous} \pressure_i^{\saturatedvapor, \rigorous} \right) \\
    \label{eq:implicitraoultslaw:modelfluid}
    F^{\modelfluid} &= \sum_{i=1}^{N_C} \left( \vapormolarfractionof{i} \pressure - \liquidmolarfractionof{i} \activitycoefficient^{\modelfluid}_i \pressure_i^{\saturatedvapor, \modelfluid} \right)
\end{align}
To analyze the difference between their solutions ($\delta \temperature = \temperature^{\rigorous} - \temperature^{\modelfluid}$ and $\delta \vapormolarfractionof{i} = \vapormolarfractionof{i}^{\rigorous} - \vapormolarfractionof{i}^{\modelfluid}$), we introduce a composite function $F^D$ that transitions from the modelfluid to the rigorous model via a parameter $t \in [0, 1]$:
\begin{equation} \label{eq:implicitraoultslawdifference}
    F^D = (1-t) \cdot F^{\modelfluid} + t \cdot F^{\rigorous} = 0.
\end{equation}
At $t=0$, this yields the modelfluid solution, and at $t=1$, the rigorous solution.
By summing over all components $i$ and using the fact that $\sum \vapormolarfractionof{i} = 1$, we can eliminate the vapor molar fractions $\vapormolarfractionof{i}$ to obtain a new implicit function, $\tilde{F}^D$, that defines the equilibrium temperature $\temperature$:
\begin{equation}
    \tilde{F}^D\inb{\temperature, t} = \pressure - \sum_{i=1}^{N_C} \left[ (1-t) \liquidmolarfractionof{i} \activitycoefficient^{\modelfluid}_i \pressure_i^{\saturatedvapor, \modelfluid} + t \liquidmolarfractionof{i} \activitycoefficient_i^{\rigorous} \pressure_i^{\saturatedvapor, \rigorous} \right] = 0.
\end{equation}
Using the Implicit Function Theorem, we can find the sensitivity of the temperature to the parameter $t$:
\begin{equation} \label{eq:partialtemperaturet}
    \frac{\partial \temperature}{\partial t} = - \left[\frac{\partial \tilde{F}^D}{\partial \temperature}\right]^{-1} \frac{\partial \tilde{F}^D}{\partial t}.
\end{equation}
A first-order Taylor expansion around $t=0$ (the modelfluid solution) gives an approximate solution for the temperature at any $t$.
Specifically, the difference between the rigorous temperature ($t=1$) and the modelfluid temperature ($t=0$) is:
\begin{equation}
    \delta \temperature = \temperature^{\rigorous} - \temperature^{\modelfluid} \approx \left. \frac{\partial \temperature}{\partial t} \right|_{t=0}.
\end{equation}
With this approximation for the temperature difference, we can then find the error in the vapor mole fraction, $\delta\vapormolarfractionof{i}$, by substituting the respective temperatures back into Raoult's Law for each model.
This provides an approximate, analytic expression for the modeling error as a function of the rigorous parameters and the system state:
\begin{equation}
    \delta\vapormolarfractionof{i} = \delta\vapormolarfractionof{i}\inb{\parameter^{\rigorous}, \liquidmolarfractions, \pressure}.
\end{equation}

\paragraph{Numeric View}
To quantify the modeling error in practice, we solved Raoult's Law for all 848 binary systems in our database using both the rigorous and modelfluid representations.
The rigorous model employed the extended Antoine equation \cite{Wilding1998} and the NRTL model \cite{Renon1968}.
All calculations were performed at a fixed state of $\liquidmolarfractionof{1} = 0.5 \:\molunit\:\molunit^{-1}$ and $\pressure = 1\:\barunit$.

\begin{figure}[h]
    \centering
    \includegraphics[width=\textwidth]{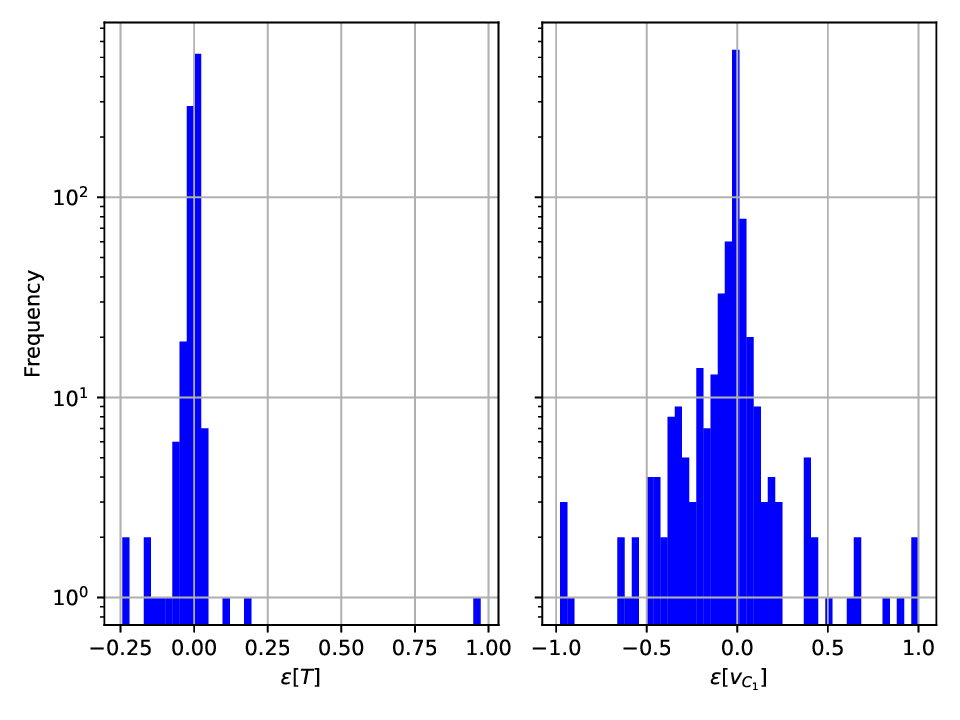}
    \caption{
        Error histogram of the modelfluid representation.
        The relative error metrics for $\varepsilon\left[\temperature\right]$ and $\varepsilon\left[\vapormolarfractionof{1}\right]$ are defined in \eqref{eq:modelerror:t} and \eqref{eq:modelerror:y}, respectively.
        The vertical axis is logarithmic to highlight the distribution of errors.
    }
    \label{fig:ModelError:Histo}
\end{figure}

\figureref{fig:ModelError:Histo} shows histograms of the relative errors for the calculated equilibrium temperature and vapor-phase mole fraction, defined as:
\begin{align} 
    \label{eq:modelerror:t}
    \varepsilon\left[\temperature\right] &= \frac{\temperature^{\modelfluid} - \temperature^{\rigorous}}{\temperature^{\rigorous}} \\
    \label{eq:modelerror:y}
    \varepsilon\left[\vapormolarfractionof{1}\right] &= \frac{\vapormolarfractionof{1}^{\modelfluid} - \vapormolarfractionof{1}^{\rigorous}}{\vapormolarfractionof{1}^{\rigorous}}.
\end{align}

The results show that for the vast majority of systems, the modeling error is small, validating the modelfluid's use as a reliable surrogate.
However, there are some outliers with large relative errors, as can be seen in \figureref{fig:ModelError:Histo}.
We find these outliers exist for two primary reasons:
a) We have used all 848 binary systems in our database, among which are some that form "artifacts" or un-physical VLE behavior.
In \figureref{fig:unphysicalVLE}, we show such a system.
Due to the large number of systems, and the fact that such artifacts may only appear at some pressure ranges, we did filter all such systems, which may affect the results shown in \figureref{fig:ModelError:Histo}.
b) The error values visualized in \figureref{fig:ModelError:Histo} are relative errors, which means that for $\liquidmolarfractionof{1} \to 0$ \eqref{eq:modelerror:y} can become large.

The few outliers with larger errors typically correspond to exotic systems where the underlying rigorous model itself may suffer from poor parameterization or numerical instability.
Since our modelfluid is derived from the rigorous VLE, its accuracy is inherently limited by the quality of the rigorous model's predictions.
One of these exotic systems is depicted in \figureref{fig:unphysicalVLE}.

\subsection{Error of Simulation} \label{sec:simulationerror}

\paragraph{Analytic View}
The second source of error stems from using a modelfluid-tailored, simplified stage-to-stage simulation algorithm, as described in \appendixref{sec:fakeenthalpysimulation}, instead of a fully rigorous one (e.g., \cite{Hoffmann2017}).
The primary simplification is the use of temperature-independent properties, which alters the stage equilibrium calculations.

For a fixed set of column specifications (feed, split ratio $\spl$, reflux ratio $\refluxratio$, etc.), the simulation solves an implicit function $\constraint_{\column}$ to find the bottom composition $\liquidmolarfractions^{(\bottom)}$ that satisfies the column-wide material balances.
We denote the rigorous and simplified conditions as:
\begin{align}
    \constraint_{\column}\inb{\liquidmolarfractions^{(\bottom)}}^{\rigorous} &= 0 \\
    \constraint_{\column}\inb{\liquidmolarfractions^{(\bottom)}}^{\modelfluid} &= 0.
\end{align}
Because the functions are different, they yield different solutions, $\liquidmolarfractions^{\bottom, \rigorous}$ and $\liquidmolarfractions^{\bottom, \modelfluid}$. The error in the bottom composition is $\delta \liquidmolarfractions^{\bottom} = \liquidmolarfractions^{\bottom, \rigorous} - \liquidmolarfractions^{\bottom, \modelfluid}$.

To derive an expression for this error, we perform a first-order Taylor expansion of the rigorous condition around the modelfluid solution:
\begin{equation}
    \constraint_{\column}\inb{\liquidmolarfractions^{\bottom, \rigorous}}^{\rigorous} = \constraint_{\column}\inb{\liquidmolarfractions^{\bottom, \modelfluid} + \delta \liquidmolarfractions^{\bottom}}^{\rigorous} \approx \constraint_{\column}\inb{\liquidmolarfractions^{\bottom, \modelfluid}}^{\rigorous} + \jacobian_{\column}^{\rigorous} \cdot \delta \liquidmolarfractions^{\bottom} = 0,
\end{equation}
where $\jacobian_{\column}^{\rigorous}$ is the Jacobian of the rigorous simulation condition evaluated at the modelfluid solution.
Rearranging gives an approximate expression for the error:
\begin{equation}
    \label{eq:simulationerror:deltaxapprox}
    \delta \liquidmolarfractions^{\bottom} \approx - \left[\jacobian_{\column}^{\rigorous}\inb{\liquidmolarfractions^{\bottom, \modelfluid}}\right]^{-1} \cdot \constraint_{\column}^{\rigorous}\inb{\liquidmolarfractions^{\bottom, \modelfluid}}.
\end{equation}
This error in the bottom composition propagates to other calculated quantities, most notably the reboiler duty, $\reboilerduty$.
The total error in reboiler duty, $\Delta \heatduty^{\reboiler}$, can be decomposed into two parts:
\begin{equation}
    \Delta \heatduty^{\reboiler} = \underbrace{\left( \heatduty^{\reboiler, \rigorous}\inb{\liquidmolarfractions^{\distillate, \rigorous}} - \heatduty^{\reboiler, \rigorous}\inb{\liquidmolarfractions^{\distillate, \modelfluid}} \right)}_{\text{Indirect Error}} + \underbrace{\left( \heatduty^{\reboiler, \rigorous}\inb{\liquidmolarfractions^{\distillate, \modelfluid}} - \heatduty^{\reboiler, \modelfluid}\inb{\liquidmolarfractions^{\distillate, \modelfluid}} \right)}_{\text{Direct Error}}.
\end{equation}
The \textit{direct error} arises from differences in the reboiler duty equations themselves, while the \textit{indirect error} is due to the simulation predicting a different distillate composition ($\liquidmolarfractions^{\distillate, \modelfluid}$ vs $\liquidmolarfractions^{\distillate, \rigorous}$).
The indirect error can be approximated using the gradient of the rigorous reboiler duty and the error in the distillate composition, $\delta\liquidmolarfractions^{\distillate}$, which is linearly related to $\delta\liquidmolarfractions^{\bottom}$.

\paragraph{Numeric View}
To assess the magnitude of the simulation error, we compared the "simplified" and "rigorous" simulation variants for the binary system of Acetone and Chloroform.
For this comparison, both variants used a Margules activity coefficient model and a simplified Antoine equation.
The "rigorous" variant included temperature-dependent DIPPR models for enthalpy \cite{Wilding1998}, while the "simplified" variant used the temperature-independent vaporization enthalpy from our modelfluid parameterization.

\begin{figure}[h]
    \centering
    \includegraphics[width=\textwidth]{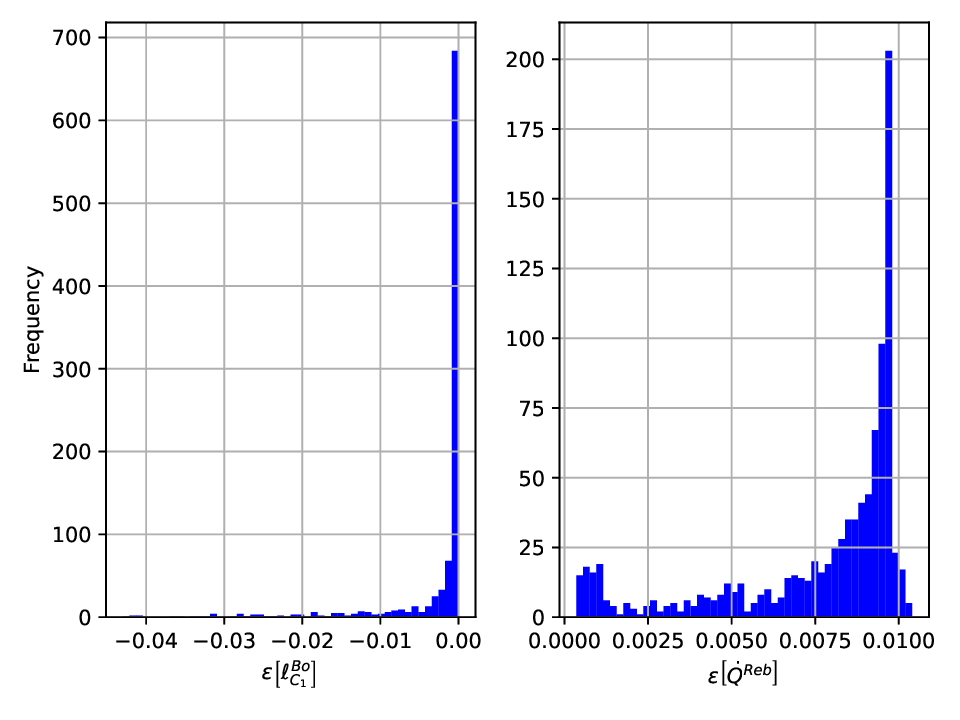}
    \caption{Simulation error histogram. The relative error metrics for $\varepsilon\left[\liquidmolarfractionof{1}^{\bottom}\right]$ and $\varepsilon\left[\reboilerduty\right]$ are defined in \eqref{eq:simerror:x} and \eqref{eq:simerror:q}, respectively.}
    \label{fig:SimError:Histo}
\end{figure}

We performed 1000 simulations with uniformly sampled operating conditions ($\mynum_{\stages}^{\belowfeed} \in [3, 30], \mynum_{\stages}^{\abovefeed} \in [3, 30], \refluxratio \in [0.1, 40], \spl \in [0.001, 0.999]$) and a fixed feed of $\liquidmolarfractionof{\Acetone}=0.5\:\molunit\:\molunit^{-1}$ at a flow of $\liquidflow^{(\feed)}=0.278\:\molunit\:s^{-1}$ and constant pressure of $1\:\barunit$.
\figureref{fig:SimError:Histo} presents histograms of the relative errors in the bottom composition and reboiler duty, defined as:
\begin{align} 
    \label{eq:simerror:x}
    \varepsilon\left[\liquidmolarfractionof{1}^{\bottom}\right] &= \frac{\liquidmolarfractionof{1}^{\bottom, \modelfluid} - \liquidmolarfractionof{1}^{\bottom, \rigorous}}{\liquidmolarfractionof{1}^{\bottom, \rigorous}} \\
    \label{eq:simerror:q}
    \varepsilon\left[\reboilerduty\right] &= \frac{\heatduty^{\reboiler, \modelfluid} - \heatduty^{\reboiler, \rigorous}}{\heatduty^{\reboiler, \rigorous}}
\end{align}
where superscript $\modelfluid$ denotes the modelfluid-tailored simulation.

The histograms demonstrate that the errors introduced by the simplified simulation are centered around zero and are generally small across a wide range of operating conditions.
This result justifies the use of our modelfluid-tailored simulation for the efficient entrainer screening and optimization tasks presented in this work.

\section{Detailed Optimization Formulation} \label{app:optimizationdetails}
This appendix provides the specific numerical values for the specifications and bounds used in the optimization problems described in Section \ref{sec:entrainerhypothetical}.
Also, additional information on the solver settings and the runtime of the different optimizations are provided.

\subsection{NQ Optimization Problem (Step 1)}

\paragraph{Specifications}
The following specifications were used for the NQ optimization problem:
\begin{itemize}
    \item \textbf{Feed Flow:} $\liquidflow_1 = 1\: kmol\: h^{-1}$.
    \item \textbf{Feed Composition:} $\liquidmolarfractions_{1} = \left[0.3, 0.4, 0.3\right]\: mol\: mol^{-1}$ for Acetone, entrainer (Benzene), and Chloroform, respectively.
    \item \textbf{Product Purity (Column 1):} Acetone is extracted from the distillate stream of column $C_1$ at a minimum purity of $0.95\: mol\: mol^{-1}$.
    \item \textbf{Product Purity (Column 2):} The entrainer is extracted from the bottom stream of column $C_2$ at a minimum purity of $0.9\: mol\: mol^{-1}$.
    \item \textbf{Product Purity (Column 3):} Chloroform is extracted from the distillate stream of column $C_3$ at a minimum purity of $0.95\: mol\: mol^{-1}$.
    \item \textbf{Minimum Product Flow:} The minimum product flow for each of the three product streams is set to $0.1\: kmol\: h^{-1}$.
\end{itemize}

\paragraph{Bounds}
The box-constraints for the optimization variables in the NQ optimization were set as follows:
\begin{itemize}
    \item \textbf{Reflux Ratio:} $\refluxratio_i \in \left[0.1, 40\right]$ for each column $i$.
    \item \textbf{Split Fraction:} $\spl_i \in \left[0.001, 0.999\right]$ for each column $i$.
    \item \textbf{Molar Fractions:} $\liquidmolarfraction_{i, \component_j} \in \left[10^{-8}, 1 - 10^{-8}\right]\: mol\: mol^{-1}$ for all components in product and recycle streams.
    \item \textbf{Stages Above Feed:} $\mynum_{\stages, \column_i}^{\abovefeed} \in \left[3, 96\right]$ for each column $i$.
    \item \textbf{Stages Below Feed:} $\mynum_{\stages, \column_i}^{\belowfeed} \in \left[3, 96\right]$ for each column $i$.
\end{itemize}
These bounds were chosen empirically to allow for sufficient exploration of the design space while cutting off excessive or numerically unstable regions. Note that the bounds on product stream compositions could be further tightened based on the minimum purity constraints to reduce the search space.

\paragraph{Numerical Details and Starting Points}
For this mixed-integer nonlinear programming (MINLP) problem, we use the \verb|MISQP| algorithm \cite{Schittkowski2014MISQP}.
For the \verb|MISQP| solver, the following settings are chosen:
\begin{itemize}
    \item Maximum number of solver iterations (\verb|MAXIT|): $\num{100}$.
    \item Maximum number of branch and bound nodes (\verb|MAXNDE|): $\num{1000}$.
    \item Termination (convergence) accuracy of the solver (\verb|ACC|): $\num{1e-6}$.
\end{itemize}
All other settings of the \verb|MISQP| solver are kept to their default value, as specified in \cite{Schittkowski2014MISQP}.
The derivatives of both the objective functions and the constraint functions w.r.t. the optimization variables are not available analytically and thus have been numerically approximated using a 3-point (mid-point) finite difference rule for the continuous variables and a 2-point rule for the discrete variables.

\subsection{Entrainer Optimization Problem (Step 2)}
\paragraph{Bounds}
For the entrainer optimization, the bounds on the number of stages were removed (as they became fixed parameters). Bounds were added for the new optimization variables corresponding to the entrainer's modelfluid features:
\begin{itemize}
    \item \textbf{Saturated Vapor Temperature:} $\saturatedvaportemperatureof{2} \in \left[300, 600\right]\: K$. This range covers realistic candidates without considering extreme high- or low-boiling liquids.
    \item \textbf{Vaporization Enthalpy:} $\vaporizationenthalpyof{2} \in \left[10, 62\right]\: kJ\: mol^{-1}$. This range was chosen to reflect realistic values and avoid unhelpful hypothetical solutions with near-zero enthalpy.
    \item \textbf{Infinite-Dilution Activity Coefficients:}
        \begin{equation}
            \activitycoefficient_i\vert_j \in \left[0.01, 8\right] \quad \forall i, j \in \{1,2,3\}, i \neq j
        \end{equation}
    \item \textbf{VLE Curve Slope at Infinite Dilution:}
        \begin{equation}
            \frac{\partial \vapormolarfractionof{i}}{\partial \liquidmolarfractionof{i}}\vert_j \in \left[0.01, 2\right] \quad \forall i, j \in \{1,2,3\}, i \neq j
        \end{equation}
\end{itemize}
The upper bounds on the activity coefficients and the derivative-based feature were chosen heuristically to ensure the optimization does not stray too far from the space of available real candidates, similar in spirit to the use of convex hulls or ellipsoids in \cite{Stavrou2023}, but less restrictive.

\paragraph{Numerical Details and Starting Points}
The entrainer optimization problem is initialized using the results from the NQ optimization.
For each Pareto-optimal point on the NQ curve, the corresponding numbers of stages are fixed as parameters for the entrainer optimization problem.
The modelfluid features of the reference entrainer (Benzene) serve as the starting point for the entrainer-specific variables $\feature_{\entrainer}$, from which the optimizer seeks to find an improvement.
We choose the \verb|NLPQLP| algorithm \cite{Schittkowski2014NLPQLP}, a sequential quadratic programming (SQP) method, due to its demonstrated robustness and efficiency in handling the highly nonlinear and non-convex problems that arise in process systems engineering.
For the \verb|NLPQLP| solver, the following settings are chosen:
\begin{itemize}
    \item Maximum number of solver iterations (\verb|MAXIT|): $\num{100}$.
    \item Termination (convergence) accuracy of the solver (\verb|ACC|): $\num{1e-10}$.
    \item Termination (convergence) accuracy of the inner quadratic problem (\verb|ACCQP|): $\num{1e-14}$.
\end{itemize}
All other settings of the \verb|NLPQLP| solver are kept to their default value, as specified in \cite{Schittkowski2014NLPQLP}.

The derivatives of both the objective functions and the constraint functions w.r.t. the optimization variables are not available analytically and thus have been numerically approximated using a 3-point (mid-point) finite difference rule.

\subsection{Runtime discussion}
This section provides further detail on the computational effort of the proposed methodology as well as the hardware it was run on.
A core ingredient of the stage-to-stage simulation algorithm is the repeated solution of extended Raoult's Law.
We use a Newton solver with a tolerance (convergence) criterion of $\num{1e-9}$ for the residual norm and a maximum (stop criterion) of $100$ iterations.
For the solution of Raoult's Law, analytical derivatives are provided to the solver.
Benchmarking studies yield that the solver takes between 4-8 iterations on average to converge.

For the remainder part of the computations, namely the solution of the overall stage-to-stage approach to the MESH equations of each distillation column, and the final optimization of the process flowsheet, numerical approximations (2-point for integer variables and 3-point rule for continuous variables, as discussed above) are employed, as no analytic derivatives are available.
This, from a computational point of view, is more expensive than a simultaneous solution approach, where all the MESH equations are used as constraints to the optimization problem, enabling their parallel computation.
However, the stage-to-stage solution approach, while more expensive, in general has better convergence properties, especially when no good starting solutions are known.
The stage-to-stage computations make up the dominant part of the entire computational effort of the proposed methodology.
In our work, we have implemented the stage-to-stage solution approach, as described in our work, using the Rust programming language.
The optimization time varies depending on the operating point and process fluid (in particular with the numbers of stages in the flowsheet -- the more stages, the more expensive is the stage-to-stage optimization).
On average, the computational effort for
\begin{itemize}
    \item Step 1: One point on the Pareto frontier is about 4 minutes.
    \item Step 2: One hypothetical entrainer solution for a fixed, given number of stages is about 15 minutes.
\end{itemize}
Step 1 and 2 are described at the beginning of this section.
The above benchmarks have been obtained on the system described in \tabref{tab:cpu_specs}, without the use of parallelization (while generally possible).

For the NQ curves computed in step 1, using a discretization of $\mynum_{\stages}^{\total}=30$ to $\mynum_{\stages}^{\total}=60$ numbers of stages in the flowsheet in combination with a comparably small number of $16$ multi start locations means about $24$ hours per Pareto frontier.
While parallelization (on a decently capable CPU) could bring this down to a couple of hours, this computational effort is necessary to validate the results from the mapping approach presented in this work.
This means the above specified computational effort for the rigorous computation of an NQ curve for the entrainer distillation flowsheet had to be paid for each of the candidates in the entrainer.

For the process fluid optimization conduced in step 2, namely the computation of the hypothetically optimal entrainer fluid for a given flowsheet configuration (distillation column stages and feed positions for each of the $30$ discretization points on the NQ curve), an approximate computational effort of 7-8 hours is necessary.
The computational cost of the mapping strategy, where the objective function for each of the candidates in the entrainer pool is simply evaluated by inserting their modelfluid features into \eqref{eq:opexprediction} is negligible ($<1$ second per candidate).

\begin{landscape}
    \begin{longtable}{@{} p{0.18\linewidth} p{0.22\linewidth} p{0.56\linewidth} @{} }
    \caption{System Processor Specifications for Computational Analysis}\label{tab:cpu_specs} \\
        \toprule
        \textbf{Feature} & \textbf{Value} & \textbf{Description} \\
    \midrule
    \endfirsthead
        \toprule
        \textbf{Feature} & \textbf{Value} & \textbf{Description} \\
    \midrule
    \endhead
    \bottomrule
    \endfoot
    \bottomrule
    \endlastfoot
        \textbf{Processor Model}         & Intel(R) Xeon(R) Gold 6248R CPU                           & Identifies the specific CPU architecture and family. \\
        \textbf{Base Clock Speed}        & \SI{3.00}{\giga\hertz}                                    & The nominal operating frequency of the CPU cores. \\
        \textbf{Architecture}            & x86\_64                                                    & Standard 64-bit instruction set architecture. \\
        \textbf{Total Logical CPUs}      & 16                                                        & The total number of processing units available to the operating system. \\
        \textbf{Physical Cores}          & 16                                                        & The number of independent processing units within the CPU. (Derived from 16 single-core virtual sockets). \\
        \textbf{Threads per Core}        & 1                                                         & Indicates that each physical core executes a single thread; Simultaneous Multi-Threading (SMT) is not enabled. \\
        \textbf{L1d Cache (total)}       & \SI{512}{\kibi\byte}                                      & Total L1 Data Cache available (\SI{32}{\kibi\byte} per core). \\
        \textbf{L1i Cache (total)}       & \SI{512}{\kibi\byte}                                      & Total L1 Instruction Cache available (\SI{32}{\kibi\byte} per core). \\
        \textbf{L2 Cache (total)}        & \SI{16}{\mebi\byte}                                       & Total L2 Cache available (\SI{1}{\mebi\byte} per core). \\
        \textbf{L3 Cache (total)}        & \SI{572}{\mebi\byte}                                      & Total L3 Cache available, typically shared among cores. Presented as a unified pool to the VM. \\
        \textbf{Vector Extensions}       & AVX, AVX2, AVX512F, AVX512DQ, AVX512BW, AVX512VL           & Advanced instruction sets (SIMD) for parallelizing certain computations. \\
    \end{longtable}
\end{landscape}

\textit{Note on Virtualization} in \tabref{tab:cpu_specs}: The system was identified as running within a VMware virtualized environment.
This context is important as virtualization can introduce minor performance overhead compared to bare-metal execution.

\section{List of Entrainer Candidates} \label{sec:availableentrainercandidates}
The entrainer candidates considered in this work were selected through a multi-stage filtering process designed to find suitable compounds for separating the Acetone+Chloroform maximum-boiling azeotrope.
The process began with a large pool of substances identified via the matrix completion method \cite{Jirasek2020}, as described in \appendixref{sec:entraineravailable}.
This list was then screened using the physical property criteria for entrainers from \cite{Stichlmair1992}.

The most significant filter, however, was a practical one: many candidates, despite having promising initial properties, led to convergence failures when simulated in the full entrainer distillation process flowsheet (\sectionref{sec:entrainerdistillation}).
These failures often arose from inconsistencies between the predicted infinite dilution activity coefficients and the available pure component property models (vapor pressure and enthalpy).
This necessary step of ensuring robust simulation resulted in the final, shorter list of viable entrainer candidates presented in Table~\ref{tab:optimizableentrainercandidates}.

\begin{longtable}{ll}

\caption{List of viable real-world entrainer candidates and their CAS Numbers that allow for successful process simulation.}
\label{tab:optimizableentrainercandidates} \\

\toprule
Entrainer & CAS Number \\
\midrule
\endfirsthead

\caption[]{(continued)} \\
\toprule
Entrainer & CAS Number \\
\midrule
\endhead

\midrule
\multicolumn{2}{r}{\textit{Continued on next page}} \\
\endfoot

\bottomrule
\endlastfoot

1,4-Butanediol & 110-63-4 \\
Diamylamine & 2050-92-2 \\
Triallylamine & 102-70-5 \\
Trans-3-Hexene & 13269-52-8 \\
Dimethyl sulfoxide & 67-68-5 \\
Methyl methacrylate & 80-62-6 \\
Benzene & 71-43-2 \\
n-Octane & 111-65-9 \\
Isoamyl alcohol & 123-51-3 \\
n-Butyl acetate & 123-86-4 \\
3-Methylheptane & 589-81-1 \\
Butyl acrylate & 141-32-2 \\
1-Methylnaphthalene & 90-12-0 \\
2-Cyanoethyl ether & 1656-48-0 \\
Formaldehyde Diethyl Acetal & 462-95-3 \\
2,4-Dimethylpentane & 108-08-7 \\
1,4-Dioxane & 123-91-1 \\
Toluene & 108-88-3 \\
Dibutylamine & 111-92-2 \\
2-Propanethiol & 75-33-2 \\
Isoquinoline & 119-65-3 \\
cis-2-Pentene & 627-20-3 \\
Ethyl tert-amyl ether & 919-94-8 \\
o-Cresol & 95-48-7 \\
Pentanoic acid & 109-52-4 \\
1,3,5-Trimethylcyclohexane & 1795-26-2 \\
1,1,2,2-Tetrabromoethane & 79-27-6 \\
epsilon-Caprolactam & 105-60-2 \\
Pyrrole & 109-97-7 \\
Thiophene & 110-02-1 \\
Hexanal & 66-25-1 \\
2-Ethyltoluene & 611-14-3 \\
Hexamethyldisiloxane & 107-46-0 \\
Trichloroethylene & 79-01-6 \\
sec-Butyl Ether & 6863-58-7 \\
2-Methyl-1-Butanol & 137-32-6 \\
2,2,4,4,6,8,8-Heptamethylnonane & 4390-04-9 \\
Furan & 110-00-9 \\
Di-n-butyl phthalate & 84-74-2 \\
Ethyl pivalate & 3938-95-2 \\

\end{longtable}

In addition to screening real-world compounds, the process fluid optimization described in \sectionref{sec:entrainersearch} identified a set of high-performing \textit{hypothetical} entrainers.
These candidates, listed in Table~\ref{tab:hypotheticalentrainercandidates}, represent optimal property sets found by the algorithm.

\begin{table}[h]
    \centering
    \begin{tabular}{lllllllll}
\toprule
$\temperature_2^{\saturatedvapor}\:/\:K$ & $\activitycoefficient_1\vert_2$ & $\activitycoefficient_2\vert_1$ & $\activitycoefficient_2\vert_3$ & $\activitycoefficient_3\vert_2$ & $\frac{\partial \vapormolarfractionof{2}}{\partial \liquidmolarfractionof{2}}\vert_1$ & $\vaporizationenthalpyof{2}\inb{\temperature\vert_2}$ & $\totalreboilerduty\:/\:kW$ & $\mynum_{\stages}^{\total}$ \\
\midrule
364.431 & 1.362 & 1.774 & 0.277 & 0.641 & 0.702 & 43013.74 & 189.65 & 35 \\
353.467 & 1.406 & 1.645 & 0.372 & 0.696 & 0.732 & 46434.8 & 170.08 & 37 \\
354.125 & 1.415 & 1.644 & 0.372 & 0.687 & 0.733 & 46466.247 & 151.7 & 38 \\
368.898 & 1.892 & 3.739 & 0.253 & 0.726 & 0.812 & 10000 & 11.14 & 39 \\
383.06 & 1.533 & 3.366 & 0.217 & 0.616 & 0.926 & 22687.785 & 12.3 & 40 \\
368.683 & 2.254 & 4.081 & 0.294 & 0.845 & 0.502 & 33201.204 & 13.09 & 41 \\
374.283 & 2.135 & 3.971 & 0.28 & 0.805 & 0.739 & 26321.812 & 11.81 & 42 \\
371.875 & 1.088 & 2.752 & 0.179 & 0.496 & 0.527 & 42387.799 & 22.18 & 43 \\
357.004 & 1.429 & 1.652 & 0.36 & 0.73 & 0.732 & 46386.122 & 75.25 & 44 \\
357.725 & 1.432 & 1.657 & 0.355 & 0.739 & 0.732 & 46330.252 & 66.76 & 45 \\
373.963 & 0.988 & 2.554 & 0.174 & 0.463 & 0.594 & 43571.29 & 20.45 & 46 \\
360.749 & 1.449 & 1.665 & 0.344 & 0.76 & 0.732 & 46291.354 & 53.56 & 48 \\
374.7 & 2.18 & 4.013 & 0.285 & 0.82 & 0.508 & 32606.568 & 11.92 & 49 \\
373.629 & 2.229 & 4.058 & 0.291 & 0.836 & 0.52 & 34624.684 & 12.08 & 50 \\
378.543 & 0.958 & 2.493 & 0.173 & 0.454 & 0.649 & 43623.271 & 16.17 & 51 \\
378.117 & 0.964 & 2.34 & 0.174 & 0.441 & 0.631 & 44736.956 & 17.23 & 52 \\
371.61 & 2.229 & 4.058 & 0.291 & 0.836 & 0.623 & 37301.423 & 11.75 & 53 \\
380.066 & 1.651 & 3.493 & 0.229 & 0.652 & 0.619 & 31448.495 & 10.84 & 54 \\
379.936 & 1.272 & 2.191 & 0.187 & 0.536 & 0.681 & 44893.872 & 15.05 & 56 \\
379.77 & 1.569 & 1.871 & 0.207 & 0.571 & 0.711 & 45609.51 & 16.07 & 57 \\
381.76 & 1.544 & 1.867 & 0.205 & 0.564 & 0.709 & 45589.931 & 15.37 & 58 \\
380.033 & 1.407 & 3.225 & 0.206 & 0.582 & 0.768 & 41347.417 & 12.09 & 59 \\
383.337 & 1.24 & 3.027 & 0.19 & 0.542 & 0.711 & 42503.955 & 12.28 & 60 \\
\bottomrule
\end{tabular}

    \caption{List of hypothetical entrainer candidates found during the process fluid optimization in \sectionref{sec:entrainersearch}.}
    \label{tab:hypotheticalentrainercandidates}
\end{table}

A key characteristic of these hypothetical entrainers is that they are all high-boiling components relative to the original mixture.
For context, the boiling points of the base components and the reference entrainer (Benzene) at 1~bar are:
\begin{itemize}
    \item Acetone: 329.04 K
    \item Chloroform: 333.22 K
    \item Benzene: 352.85 K
\end{itemize}
As shown in Table~\ref{tab:hypotheticalentrainercandidates}, all hypothetical entrainers possess boiling points significantly higher than that of Benzene, highlighting a key physical trait for successful entrainers in this specific separation task.

We note that the list of viable real candidates could likely be expanded.
The use of more sophisticated, temperature-dependent models for predicting infinite dilution activity coefficients, such as those by \cite{Damay2021}, \cite{Rittig2023}, or \cite{Specht2024}, might resolve some of the simulation issues encountered.
However, for the scope of this case study, the current set of candidates was deemed sufficient.

\section{Additional data and figures}
\begin{figure}[h]
    \centering
    \includegraphics[width=\textwidth]{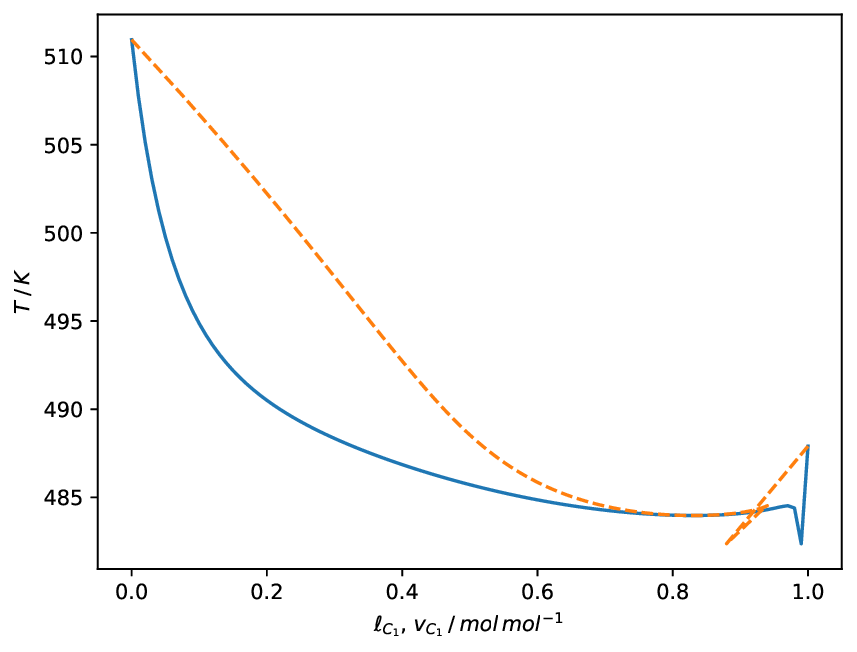}
    \caption{Vapor liquid equilibrium for the system Triethylphosphat + Bicyclohexyl at $\pressure = 1\,\barunit$ using the parameters in our database. This VLE is used as an example of an un-physical VLE.}
    \label{fig:unphysicalVLE}
\end{figure}

The vapor liquid equilibrium shown in \figureref{fig:unphysicalVLE} is an example of an un-physical VLE that we found in our mixture database. Those artifacts are rare which makes it difficult to filter them out. Since some of these systems are contained in the results shown in the numeric analysis of \appendixref{sec:modelingerror} and \appendixref{sec:simulationerror}, resulting in outliers in those discussions, we show how such a VLE curve looks like in \figureref{fig:unphysicalVLE}, since we refer to it in the manuscript.

\clearpage


\end{document}